\documentclass[preprint,review,11pt]{elsarticle}

\usepackage{amssymb}
\usepackage{amsmath}
\usepackage{subfigure}
\usepackage{array}
\usepackage{multirow}
\usepackage{threeparttable}
\biboptions{numbers,sort&compress}
\usepackage{float}
\usepackage{geometry}   
\journal{Signal Processing}

\begin{document}

\begin{frontmatter}



\title{Multi-dimensional Graph Linear Canonical Transform}


\author[1]{Na Li}
\author[1,2,3]{Zhichao Zhang\corref{cor1}}
\ead{zzc910731@163.com}
\author[4]{Jie Han}
\author[1]{Yunjie Chen}
\author[1]{Chunzheng Cao}
\address[1]{School of Mathematics and Statistics, Center for Applied Mathematics of Jiangsu Province, and Jiangsu International Joint Laboratory on System Modeling and Data Analysis, Nanjing University of Information Science and Technology, Nanjing 210044, China}
\address[2]{Key Laboratory of System Control and Information Processing, Ministry of Education, Shanghai 200240, China}
\address[3]{Key Laboratory of Computational Science and Application of Hainan Province, Haikou 571158, China}
\address[4]{School of Remote Sensing and Geomatics Engineering, Nanjing University of Information Science and Technology, Nanjing 210044, China}

\cortext[cor1]{Corresponding author; Tel: +86-13376073017.}
\tnotetext[mytitlenote]{This work was supported by the National Natural Science Foundation of China under Grant No 61901223, the Jiangsu Planned Projects for Postdoctoral Research Funds under Grant No 2021K205B, the Foundation of Key Laboratory of System Control and Information Processing, Ministry of Education under Grant No Scip20240121, and the Foundation of Key Laboratory of Computational Science and Application of Hainan Province under Grant No JSKX202401.}


\begin{abstract}
Many multi-dimensional (M-D) graph signals appear in the real world, such as digital images, sensor network measurements and temperature records from weather observation stations. It is a key challenge to design a transform method for processing these graph M-D signals in the linear canonical transform domain. This paper proposes the two-dimensional graph linear canonical transform based on the central discrete dilated Hermite function (2-D CDDHFs-GLCT) and the two-dimensional graph linear canonical transform based on chirp multiplication-chirp convolution-chirp multiplication decomposition (2-D CM-CC-CM-GLCT). Then, extending 2-D CDDHFs-GLCT and 2-D CM-CC-CM-GLCT to M-D CDDHFs-GLCT and M-D CM-CC-CM-GLCT. In terms of the computational complexity, additivity and reversibility, M-D CDDHFs-GLCT and M-D CM-CC-CM-GLCT are compared. Theoretical analysis shows that the computational complexity of M-D CM-CC-CM-GLCT algorithm is obviously reduced. Simulation results indicate that M-D CM-CC-CM-GLCT achieves comparable additivity to M-D CDDHFs-GLCT, while M-D CM-CC-CM-GLCT exhibits better reversibility. Finally, M-D GLCT is applied to data compression to show its application advantages. The experimental results reflect the superiority of M-D GLCT in the algorithm design and implementation of data compression.

\end{abstract}


\end{frontmatter}


\section{Introduction}
\label{}
In traditional signal processing, signals are typically represented in regular format. However, with the rapid development of information technology, signals encountered in practical problems frequently involve complex high-dimensional data. These signals has an irregular topological structure distinct from the conventional time-space domain, so requires new processing methods. Therefore, the emergence of graph signal processing technology occurred \cite{Shuman2012TheEF,Sandryhaila2012DiscreteSP,Sandryhaila2014BigDA,Alikaifolu2024WienerFI}. Graph signal processing extends conventional discrete signal processing to graph signals with topological structures, offering an effective approach for handling data with intricate structures. This technology finds widespread application in radar surveillance, biomedicine, image processing, and other domains. In order to study graph signals, it is necessary to extend the concepts from traditional signal processing to graph signal domain. This includes fundamental concepts such as graph signal transform, convolution, filtering, shifting, and modulation \cite{Domingos2020GraphFT,Yang2019GraphFT,Jiang2019NonsubsampledGF,Ramakrishna2020AUG,Xiao2021DistributedNP}, which collectively establish a complete theoretical system for graph signal processing. 

There are two main construction frameworks for graph signal processing, which are based on spectral method and algebraic method respectively. The first framework is based on the construction of Laplace matrix, and intuitively analyzes the graph structure based on spectrogram theory. Specifically, the graph Fourier transform (GFT) is defined by the eigenvector of Laplacian matrix, and the corresponding spectrum are represented by eigenvalues \cite{Shuman2012TheEF}. Since the Laplacian matrix must be symmetric and positive semidefinite, this method is exclusively applicable to undirected graphs. The second framework is derived from the discrete signal processing on graphs, which analyzes the characteristics of graph structure signals according to algebraic methods. Specifically, GFT is defined by using the eigenvector of adjacency matrix, and the corresponding spectrum are represented by eigenvalues \cite{Sandryhaila2014BigDA}. Since the adjacency matrix may not be symmetric, this method is applicable to any graph. Similar to the classical Fourier transform, GFT may not sufficiently address certain mathematical challenges in practical applications. Hence, new transform methods have been introduced. To address the limitation of GFT in accurately representing the spectrum of multi-dimensional graph signals defined in Cartesian product graphs, the multi-dimensional GFT (M-D GFT) \cite{Kurokawa2017MultidimensionalGF,Natali2020ForecastingMP,Varma2018SAMPLINGTF} has been proposed. Using the algebraic property of Cartesian product, M-D GFT rearranges the one-dimensional(1-D) spectrum obtained by GFT into M-D frequency domain, where each dimension represents the directional frequency along each factor graph. To enhance the extraction of local graph signal features, the fractional Fourier transform (FRFT) has been introduced into graph signal processing \cite{Wang2017TheFF,Wu2019FractionalSG,Yan2021WindowedFF,Kartal2022JointTF}. The incorporation of the rotation angle parameter allows signal representation in the fractional domain situated between the time and frequency domains. The optimal representation of signals in fractional domains is achieved by analyzing the signals across all fractional domains, encompassing both the time and frequency domains. However, akin to GFT, the graph FRFT (GFRFT) encounters the same challenge in representing the spectrum of M-D graph signals, ignoring the product structure in the graph. Particularly for graphs with numerous nodes, signal processing operations need more computational complexity. Analogous to the generalization from GFT to M-D GFT, proposed the definition of the M-D GFRFT \cite{Yan2021MultidimensionalGF}. This definition avoids the binary spectrum of one-dimensional GFRFT(1-D GFRFT), shows the directional characteristics of M-D graph signals and reduces the computational complexity of graph fractional domain.

Linear canonical transform (LCT) expands rotation to affine transformation, thereby extending the Fourier transform (FT) and FRFT as a more expansive parameterized linear integral transform \cite{Wei2021SparseDL,Urynbassarova2021DiscreteQL,Goel2022ApplicationsOT,Ciobanu2021ModelingCC}. Compared with FT and FRFT, LCT introduces three independent parameters, enabling non-band-limited signals in traditional FT and FRFT domains to maybe become band-limited signals in the LCT domain under specific parameter settings. Unlike the rotational properties of FRFT in the time-frequency plane, LCT exhibits affine transformation characteristics that encompass compression and stretching operations. Therefore, it is necessary to extend LCT to graph signals processing. Zhang et al. \cite{Zhang2022DiscreteLC} introduced a definition of the graph LCT based on central discrete dilated Hermite function (CDDHFs-GLCT), achieved through a combination of graph chirp multiplication(CM) \cite{Pei2002EigenfunctionsOL,Satyan2010ChirpMB}, graph scaling transform, and GFRFT. While this discrete algorithm exhibits favorable properties, it is associated with high computational complexity, thereby posing challenges for practical application. To overcome this shortcoming, Li et al. \cite{ZhangGraphLC} proposed a GLCT based on chirp
multiplication-chirp convolution-chirp multiplication decomposition (CM-CC-CM-GLCT), which irrelevants to the sampling periods and without oversampling operation. Given that the GLCT offers greater flexibility and freedom compared to the GFRFT, and can achieve better signal representation, it is necessary to generalize the M-D GFRFT to the multi-dimensional graph linear canonical transform (M-D GLCT). Besides, when studying the M-D graph signals on the Cartesian product graphs, GLCT still has the possibility of binary spectrum and ignores the directional information. In order to make up for this deficiency, this paper proposes two kinds of M-D GLCT applied to M-D graph signals, which are based on CDDHFs and CM-CC-CM decomposition respectively. On the one hand, the M-D GLCT rearranges the 1-D spectrum obtained by GLCT into the M-D frequency domain and provides the M-D spectrum of signals. In addition to the frequency characteristics provided by traditional GLCT, the proposed M-D GLCT also include directional information. This shows that M-D GLCT provides more in-depth frequency analysis of multi-dimensional graph signals than GLCT. On the other hand, compared with M-D GFT and M-D GFRFT, M-D GLCT provides more freedom and has enough flexibility in processing M-D graph signals, showing its advantages in signal denoising, filtering and other applications. Therefore, conducting an in-depth study of M-D GLCT holds significant research value. The main contributions of this paper are summarized as follows:  \\
$\bullet$ The paper proposes the 2-D CDDHFs-GLCT and the 2-D CM-CC-CM-GLCT, which are both irrelevant to the sampling periods and without oversampling operation. Extending it to multi-dimensional, M-D CDDHFs-GLCT and M-D CM-CC-CM-GLCT are defined. \\
$\bullet$ In terms of computational complexity, M-D CM-CC-CM-GLCT exhibits lower computational complexity than M-D CDDHFs-GLCT. It has higher efficiency and performance in practical applications. \\
$\bullet$ In terms of additivity and reversibility, M-D CM-CC-CM-GLCT can achieve similar additivity compared with M-D CDDHFs-GLCT. Most importantly, M-D CM-CC-CM-GLCT has better reversibility, which offers an auxiliary thinking in some applications such as coding and decoding. \\
$\bullet$ M-D GLCT is employed for data compression to demonstrate its advantages. Experimental results indicate that M-D GLCT achieves closer fidelity to the original data compared to M-D GFRFT at equivalent compression ratios, with smaller errors.

The structure of this paper is as follows. In Section 2, some basic two-dimensional (2-D) graph operations are briefly summarized, including 2-D GFT, 2-D GFRFT, 2-D graph CM and 2-D graph scaling transform. In Section 3, defines the 2-D GLCT based on CDDHFs decomposition (2-D CDDHFs-GLCT) and the 2-D GLCT based on CM-CC-CM decomposition (2-D CM-CC-CM-GLCT). In Section 4,extends 2-D CDDHFs-GLCT and 2-D CM-CC-CM-GLCT to M-D CDDHFs-GLCT and M-D CM-CC-CM-GLCT. In Section 5, the M-D CDDHFs-GLCT and the M-D CM-CC-CM-GLCT are compared in terms of computational complexity, additivity and reversibility. In Section 6, M-D GLCT is applied to data compression to illustrate its application advantages. In Section 7, the paper is summarized.


\section{Basic Two-dimensional Graph Operations}
\label{}

Based on the decomposition of CDDHFs and CM-CC-CM, this paper proposes two kinds of 2-D GLCT respectively. In this section, we introduce some basic 2-D graph operations used in 2-D GLCT.

\subsection{Cartesian product graph}

Consider a graph $G\left ( V,E,\mathbf{A} \right )$, where $V$ represents the set of $N$ nodes in the graph, $E$ is the set of edges, and $\mathbf{A}$ is the weighted adjacency matrix. For any two different nodes $v_{i},v_{j}\in V$, $\mathbf{A}\left ( i,j \right )=\omega _{ij} $, with $\omega _{ij} $ represents the connection weight between node $v_{i}$ and $v_{j}$. 

For two graphs $G_{1} \left ( V_{1} ,E_{1} ,\mathbf{A}_{1}  \right )$ and $G_{2} \left ( V_{2} ,E_{2} ,\mathbf{A}_{2}  \right )$, $G_{1}\Box G_{2}$ represents a Cartesian product graph \cite{Imrich2008TopicsIG} with vertex set $V_{1}\times  V_{2}$, where $V_{1}=\left \{ 0,\dots ,N_{1}-1  \right \}$ and $V_{2}=\left \{ 0,\dots ,N_{2}-1  \right \}$. The edge set $E$ of Cartesian product graph $G_{1}\Box G_{2}$ satisfies 
\begin{equation}
    \left \{ \left ( i_{1},i_{2}   \right ),\left ( j_{1},j_{2}   \right )   \right \} \in E \Longleftrightarrow \left [ \left \{ i_{1},j_{1}  \right \}\in E_{1},i_{2}=j_{2}    \right ] \rm{or} \left [ i_{1}=j_{1},\left \{ i_{2},j_{2}  \right \}\in E_{2} \right ]
.\end{equation}
For $n = 1, 2$, suppose that a factor graph $G_{n}$ have the degree matrix $\mathbf{D}_{n}$ and the Laplacian matrix $\mathbf{L}_{n}$. The adjacency matrix, the degree matrix and the Laplacian matrix of $G_{1}\Box G_{2}$ can be expressed by its factor graph as $\mathbf{A}_{1}\oplus \mathbf{A}_{2}$, $\mathbf{D}_{1}\oplus \mathbf{D}_{2}$ and $\mathbf{L}_{1}\oplus \mathbf{L}_{2}$, respectively. Here, the operator $\oplus$ is a Kronecker sum  \cite{Kadambari2019LearningPG} defined by $\mathbf{A}\oplus \mathbf{B}=\mathbf{A}\otimes \mathbf{I}_{n}+\mathbf{I}_{m}\otimes \mathbf{B}$ for matrix $\mathbf{A}\in \mathbb{R}^{m\times m}$ and $\mathbf{B}\in \mathbb{R}^{n\times n}$, where $\mathbf{I}_{n}$ is the identity matrix of size $n$. 

\subsection{Two-dimensional graph Fourier transform}
Most approaches essentially define the GFT by decomposing of a general graph shift operator (GSO) $\mathbf{Z}_{n}$. We consider the graph Laplacian $\mathbf{Z}_{n}=\mathbf{L}_{n}=\mathbf{D}_{n}-\mathbf{A}_{n}$ and the Adjacency $\mathbf{Z}_{n}=\mathbf{A}_{n}$. The Jordan decomposition of $\mathbf{Z}_{n}$ is obtained, $\mathbf{Z}_{n}=\mathbf{V}_{\mathbf{Z}_{n} }\mathbf{J}_{\mathbf{Z}_{n} }\mathbf{V}_{\mathbf{Z}_{n} }^{-1}$, so that we define the GFT matrix as $\mathbf{F}_{G_{n} }=\mathbf{V}_{\mathbf{Z}_{n} }^{-1}$ . For the most commonly used undirected real weighted graphs, there is $\mathbf{Z}_{n}=\mathbf{V}_{\mathbf{Z}_{n} }\mathbf{\Lambda}_{\mathbf{Z}_{n} }\mathbf{V}_{\mathbf{Z}_{n} }^{-\mathrm{H}}$, where $\mathbf{\Lambda}_{\mathbf{Z}_{n} }$ be a diagonal matrix composed of eigenvalues $\left \{  \lambda _{i} \right \}$, $\mathbf{V}_{\mathbf{Z}_{n} }$ be a matrix composed of eigenvectors $\left \{  \mathbf{u}_{i} \right \}$, and $\mathbf{V}_{\mathbf{Z}_{n} }^{-\mathrm{H}}$ be the inverse matrix of $\mathbf{V}_{\mathbf{Z}_{n} }$. Eigenvalues and eigenvectors contain the information of graph structure signals. The larger the eigenvalues, the more high-frequency information contained in the corresponding feature components, and the more violent the fluctuations. On the contrary, the eigenvector with smaller eigenvalues corresponds to the low-frequency eigenvector, which describes the smooth characteristics of graph signals and has less fluctuation. For the convenience of discussion, we mainly discuss undirected real weighted graphs in this paper.

Definition 1. The 1-D GFT of signal $\mathbf{x}$ on a graph $G_{n}$ is defined by decomposing of the GSO $\mathbf{Z}_{n}$ as
\begin{equation}
O_{\rm{1-D \ GFT}}\left ( \mathbf{x} \right )=\mathbf{\mathbf{F}_{G_{n} }x}=\mathbf{V}_{\mathbf{Z}_{n} }^{-\mathrm{H}}\mathbf{x},  
\end{equation}
where $\mathbf{\mathbf{F}_{G_{n} }}=\mathbf{V}_{\mathbf{Z}_{n} }^{-\mathrm{H}} $ is the graph Fourier transform matrix and where $O_{\rm{1-D \ GFT}}$ denotes the 1-D GFT operator. $\hat{\mathbf{x}}=\left [ \hat{x }_{0},\hat{x }_{1},\dots ,\hat{x }_{N-1}\right ]^{\rm{T}}$.

Let's consider 2-D GFT based on GSOs on Cartesian product graph $G_{1}\Box G_{2}$. The characteristic decomposition of the GSO matrix $\mathbf{Z}_{n}$ of graph $G_{n}$ is $\mathbf{Z}_{n}=\mathbf{V}_{\mathbf{Z}_{n} }\mathbf{\Lambda}_{\mathbf{Z}_{n} }\mathbf{V}_{\mathbf{Z}_{n} }^{-\rm{H}}$, $n=1,2$.

Definition 2. The 2-D GFT of signal $x:V_{1}\times V_{2}\longrightarrow \mathbb{R}$ on a Cartesian product graph $G_{1}\Box G_{2}$ is defined by decomposing of GSOs $\mathbf{Z}_{n}$ as
\begin{equation}
\hat{x}\left ( \lambda _{k_{1} }^{\left ( 1 \right ) },\lambda _{k_{2} }^{\left ( 2 \right ) }  \right )=\sum_{i_{1} =0}^{N_{1}-1}\sum_{i_{2} =0}^{N_{2}-1}\mathbf{V}_{\mathbf{Z}_{1} }^{\ast }\left ( i_{1},  \lambda _{k_{1} }^{\left ( 1 \right ) }\right )\mathbf{V}_{\mathbf{Z}_{2} }^{\ast }\left ( i_{2},  \lambda _{k_{2} }^{\left ( 2 \right ) }\right )x\left ( i_{1},i_{2}   \right )  , 
\end{equation}
for $k_{1}=0,\dots ,N_{1} -1$ and $k_{2}=0,\dots ,N_{2} -1$, where $\ast$ denotes complex conjugate. And its inverse is defined by
\begin{equation}
x\left ( i_{1},i_{2}   \right )=\sum_{k_{1} =0}^{N_{1}-1}\sum_{k_{2} =0}^{N_{2}-1}\mathbf{V}_{\mathbf{Z}_{1} }\left ( i_{1},  \lambda _{k_{1} }^{\left ( 1 \right ) }\right )\mathbf{V}_{\mathbf{Z}_{2} }\left ( i_{2},  \lambda _{k_{2} }^{\left ( 2 \right ) }\right )\hat{x}\left ( \lambda _{k_{1} }^{\left ( 1 \right ) },\lambda _{k_{2} }^{\left ( 2 \right ) }  \right )   ,
\end{equation}
for $i_{1}=0,\dots ,N_{1} -1$ and $i_{2}=0,\dots ,N_{2} -1$.

Note that the 2-D GFT is represented as a chain of matrixmatrix multiplications:
\begin{equation}
    \hat{\mathbf{X}}=\mathbf{F}_{\mathbf{G}_{1}} \mathbf{X}\left ( \mathbf{F}_{\mathbf{G}_{2} }  \right )^{\rm{T}},
    \end{equation}
where $\mathbf{X}=\left [ x\left ( i_{1},i_{2}   \right ) \right ]_{N_{1}\times N_{2} }$, $\hat{\mathbf{X}}=\left [ \hat{x}\left ( \lambda _{k_{1} }^{\left ( 1 \right ) },\lambda _{k_{2} }^{\left ( 2 \right ) }  \right ) \right ]_{N_{1}\times N_{2} }$. Similarly, its inverse can be represented as
\begin{equation}
    \mathbf{X}=\mathbf{F}_{\mathbf{G}_{1}}^{-1} \hat{\mathbf{X}}\left ( \mathbf{F}_{\mathbf{G}_{2}}^{-1}  \right )^{\rm{T}}.
\end{equation}
The Vector formulation of 2-D GFT is represented as
\begin{equation}
    O_{\rm{2-D \ GFT}}\left ( \mathbf{x} \right )=\left ( \mathbf{F}_{\mathbf{G}_{2}}\otimes \mathbf{F}_{\mathbf{G}_{1}} \right )\mathbf{x},
\end{equation}
\begin{equation}
    O_{\rm{2-D \ GFT}}^{-1}\left ( \hat{\mathbf{x}} \right )=\left ( \mathbf{F}_{\mathbf{G}_{2}}^{-1} \otimes \mathbf{F}_{\mathbf{G}_{1}}^{-1} \right )\hat{\mathbf{x}} ,
\end{equation}
where $\mathbf{x}=vec\left ( \mathbf{X} \right )$, $\mathbf{\hat{x} }=vec\left ( \mathbf{\hat{X} } \right )$. $O_{\rm{2-D \ GFT}}$ denotes the 2-D GFT operator and $O_{\rm{2-D \ GFT}}^{-1}$ denotes the 2-D IGFT operator.

\subsection{ Two-dimensional graph fractional Fourier transform}

The Jordan decomposition of graph Fourier transform matrix $\mathbf{F}_{G_{n} }$ is obtained, $\mathbf{F}_{G_{n} }= \mathbf{V}_{\mathbf{F}_{n} }\mathbf{J}_{\mathbf{F}_{n} }\mathbf{V}_{\mathbf{F}_{n} }^{-1}$.

Definition 3. The 1-D GFRFT of signal $\mathbf{x}$ on a graph $G_{n}$ is defined by decomposing of the GSO $\mathbf{Z}_{n}$ as
\begin{equation}
O_{\rm{1-D \ GFRFT}}^{\alpha  }\left ( \mathbf{x} \right )=\mathbf{F}_{G_{n} }^{\alpha }\mathbf{x}=\mathbf{V}_{\mathbf{F}_{n} }\mathbf{J}_{\mathbf{F}_{n} }^{\alpha }\mathbf{V}_{\mathbf{F}_{n} }^{-1}\mathbf{x},
\end{equation}
where $\alpha \in \left [ 0,1 \right ]$ and $O_{\rm{1-D \ GFRFT}}^{\alpha }$ denotes the 1-D GFRFT operator.

Definition 4. The 2-D GFRFT of signal $x:V_{1}\times V_{2}\longrightarrow \mathbb{R}$ on a Cartesian product graph $G_{1}\Box G_{2}$ is defined by decomposing of GSOs $\mathbf{Z}_{1},\mathbf{Z}_{2}$ as
\begin{equation}
\begin{split}
O_{\rm{2-D \ GFRFT}}^{\alpha}\left ( \mathbf{x} \right )= & \mathbf{\left ( F_{G_{2}}\otimes F_{G_{1}} \right )^{\alpha } x} \\
= &\left ( \mathbf{V}_{\mathbf{F}_{2} }\otimes \mathbf{V}_{\mathbf{F}_{1} } \right ) \left ( \mathbf{J}_{\mathbf{F}_{2} }\otimes \mathbf{J}_{\mathbf{F}_{1} } \right ) ^{\alpha }\left ( \mathbf{V}_{\mathbf{F}_{2} }^{-1}\otimes \mathbf{V}_{\mathbf{F}_{1} }^{-1} \right ) \mathbf{x},
\end{split}
\end{equation}
where $\mathbf{X}=\left [ x\left ( i_{1},i_{2}   \right ) \right ]_{N_{1}\times N_{2} }$, $\hat{\mathbf{X}}=\left [ \hat{x}\left ( \lambda _{k_{1} }^{\left ( 1 \right ) },\lambda _{k_{2} }^{\left ( 2 \right ) }  \right ) \right ]_{N_{1}\times N_{2} }$ and $\mathbf{x}=vec\left ( \mathbf{X} \right )$, $\mathbf{\hat{x} }=vec\left ( \mathbf{\hat{X} } \right )$. $O_{\rm{2-D \ GFRFT}}^{\alpha }$ denotes the 2-D GFRFT operator.

\subsection{ Two-dimensional graph Chirp Multiplication }
In the field of graph signals, there are few theoretical documents defining graph chirp signals. Therefore, Zhang et al. \cite{Zhang2022DiscreteLC} give the graph CM matrix is defined in terms
of the chrip Fourier transform on graphs. The CM is performed on the GSO $\mathbf{Z}_{n}$, as
\begin{equation}
        \left ( \mathbf{F}_{G_{n} }  \right ) ^{\xi }=\mathbf{V}_{\mathbf{F}_{n} } \left ( \mathbf{J}_{\mathbf{F}_{n}} \right )  ^{\xi } \mathbf{V}_{\mathbf{F}_{n}}^{-1} .
\end{equation}

Definition 5. The 1-D graph CM of signal $\mathbf{x}$ on a graph $G_{n}$ is defined by decomposing of the GSO $\mathbf{Z}_{n}$ as
\begin{equation}
O_{\rm{1-D \ GCM}}^{\xi }\left ( \mathbf{x} \right )=\left ( \mathbf{J}_{\mathbf{F}_{n}} \right )  ^{\xi }\mathbf{x},
\end{equation}
where $\xi$ is the graph CM parameter and $\left ( \mathbf{J}_{\mathbf{F}_{n}} \right ) ^{\xi }$ is the eigenvalue of the matrix $\mathbf{Z}_{n}$ after CM. $O_{\rm{1-D \ GCM}}^{\xi }$ denotes the 2-D graph CM operator.

As can be seen from the above definition, chirp Fourier transform is closely related to fractional Fourier transform. Therefore, we give the definition of 2-D graph CM according to the definition of 2-D GFRFT. 

The CM is performed on GSOs $\mathbf{Z}_{1},\mathbf{Z}_{2}$, as
\begin{equation}
    \left ( \mathbf{F}_{\mathbf{G}_{2}}\otimes \mathbf{F}_{\mathbf{G}_{1}} \right )^{\xi } \ 
=  \left ( \mathbf{V}_{\mathbf{F}_{2} }\otimes \mathbf{V}_{\mathbf{F}_{1} } \right ) \left ( \mathbf{J}_{\mathbf{F}_{2} }\otimes \mathbf{J}_{\mathbf{F}_{1} } \right ) ^{\xi }\left ( \mathbf{V}_{\mathbf{F}_{2} }^{-1}\otimes \mathbf{V}_{\mathbf{F}_{1} }^{-1} \right ).
\end{equation}

Definition 6. The 2-D graph CM of signal $x:V_{1}\times V_{2}\longrightarrow \mathbb{R}$ on a Cartesian product graph $G_{1}\Box G_{2}$ is defined by decomposing of GSOs $\mathbf{Z}_{1},\mathbf{Z}_{2}$ as
\begin{equation}
\begin{split}
O_{\rm{2-D \ GCM}}^{\xi}\left ( \mathbf{x} \right )= 
 \left( \mathbf{J}_{\mathbf{F}_{2} }\otimes \mathbf{J}_{\mathbf{F}_{1} } \right ) ^{\xi } \mathbf{x},
\end{split}
\end{equation}
where $\mathbf{X}=\left [ x\left ( i_{1},i_{2}   \right ) \right ]_{N_{1}\times N_{2} }$, $\hat{\mathbf{X}}=\left [ \hat{x}\left ( \lambda _{k_{1} }^{\left ( 1 \right ) },\lambda _{k_{2} }^{\left ( 2 \right ) }  \right ) \right ]_{N_{1}\times N_{2} }$ and $\mathbf{x}=vec\left ( \mathbf{X} \right )$, $\mathbf{\hat{x} }=vec\left ( \mathbf{\hat{X} } \right )$.$O_{\rm{2-D \ GCM}}^{\xi}$ denotes the 2-D graph CM operator.

\subsection{Two-dimensional graph scaling transform}
The concept of shift is extended to general graph signals, in which the relational dependence between data is represented by arbitrary graphs. It does this by replacing the samples on nodes with weighted linear combinations of signal samples on adjacent nodes. Therefore, the transformation result of the graph shift is given by the product of the input signal and the relation matrix of the graph. There are many other possible shift operators, including adjacent weighted versions and so on. In this paper, we only consider the scale GSO $\mathbf{S}=\frac{1}{\sigma }\mathbf{Z}_{n}$ as the graph shift operator. Then the spectrum of $\mathbf{S}$ is decomposed to obtain $\mathbf{S}=\mathbf{V_{S}J_{S}V}_{\mathbf{S}}^{-1}$.

Definition 7. The 1-D graph scaling transform of signal $\mathbf{x}$ on a graph $G_{n}$ is defined by decomposing of the GSO $\mathbf{Z}_{n}$ as
\begin{equation}
    O_{\rm{1-D \ GS}}^{\sigma}\left ( \mathbf{x} \right )=\mathbf{Sx}=\mathbf{V_{S}J_{S}V}_{\mathbf{S}}^{-1} \mathbf{x},
\end{equation}
where $\sigma$ is the scaling transform factor and $O_{\rm{1-D \ GS}}^{\sigma}$ denotes the 1-D graph scaling transform operator. The definition of 1-D graph scaling transformation mentioned above is extended to 2-D graphs. 

Definition 8. The 2-D graph scaling transform of signal $x:V_{1}\times V_{2}\longrightarrow \mathbb{R}$ on a Cartesian product graph $G_{1}\Box G_{2}$ is defined by decomposing of GSOs $\mathbf{Z}_{1},\mathbf{Z}_{2}$ as
\begin{equation}
O_{\rm{2-D \ GS}}^{\sigma}\left ( \mathbf{x} \right )= 
= \left ( \mathbf{V}_{{\mathbf{S}_{2} }}\otimes \mathbf{V}_{{\mathbf{S}_{1} }} \right ) \left ( \mathbf{J}_{\mathbf{S}_{2}}\otimes \mathbf{I}_{N_{1} } + \mathbf{I}_{N_{2} }\otimes \mathbf{J}_{\mathbf{S}_{1}} \right ) \left ( {\mathbf{V}_{\mathbf{S}_{2} }^{-1} }\otimes {\mathbf{V}_{\mathbf{S}_{1} }^{-1} } \right )  \mathbf{x},
\end{equation}
where $\sigma$ is the scaling transform factor and $\mathbf{S}_{1}=\frac{1}{\sigma }\mathbf{Z}_{1}$, $\mathbf{S}_{2}=\frac{1}{\sigma }\mathbf{Z}_{2}$. The graph shift operator of graph $G_{1}\Box G_{2}$ is $\mathbf{S}=\mathbf{S}_{1}\oplus \mathbf{S}_{2}$. And $\mathbf{X}=\left [ x\left ( i_{1},i_{2}   \right ) \right ]_{N_{1}\times N_{2} }$, $\hat{\mathbf{X}}=\left [ \hat{x}\left ( \lambda _{k_{1} }^{\left ( 1 \right ) },\lambda _{k_{2} }^{\left ( 2 \right ) }  \right ) \right ]_{N_{1}\times N_{2} }$ and $\mathbf{x}=vec\left ( \mathbf{X} \right )$, $\mathbf{\hat{x} }=vec\left ( \mathbf{\hat{X} } \right )$. $O_{\rm{2-D \ GS}}^{\sigma}$ denotes the 2-D graph scaling transform operator.

\section{ Two-dimensional graph linear canonical transform }
2-D GLCT is directly extended to Cartesian product graphs through 1-D GLCT. In this section, the implementation algorithms of 2-D CDDHFs-GLCT and 2-D CM-CC-CM-GLCT are introduced. These two methods are based on the additivity of 2-D GLCT and the decomposition of parameter matrix, in which the parameter matrix is decomposed into three or five matrices. These basic discrete operations have been defined above.

\subsection{2-D CDDHFs-GLCT}
In order to further study graph signals and FRFT, the GFRFT is extended to the field of 2-D graph signals, which is similar to the extension of 1-D GFT to 2-D GFT. The parameter matrix $\mathbf{M}$ of 2-D GLCT is $4\times 4$ transform matrix, which is called ABCD matrix and also denoted by $\left ( \mathbf{A,B;C,D} \right )$ in this paper. The matrices $\mathbf{A,B,C,D}$ are all $2\times 2$ quantity matrices, that is $\mathbf{A}= \left ( \begin{matrix}a & 0\\0 &a\end{matrix} \right )$, $\mathbf{B}= \left ( \begin{matrix}b & 0\\0 &b\end{matrix} \right )$, $\mathbf{C}= \left ( \begin{matrix}c & 0\\0 &c\end{matrix} \right )$, $\mathbf{D}= \left ( \begin{matrix}d & 0\\0 &d\end{matrix} \right )$. The parameters of 2-D GLCT are the same as those of 1-D GLCT, satisfies $ad-bc=1$. 2-D GFT, 2-D GFRFT, 2-D graph CM and 2-D graph scaling transform are special cases of 2-D GLCT.

The parameter matrix of 2-D GFT is given by the following formula
\begin{equation}
    \mathbf{M}=\left ( \begin{matrix}\mathbf{A} & \mathbf{B}\\\mathbf{C} &\mathbf{D}\end{matrix} \right )=\left ( \begin{matrix}\mathbf{0} & \mathbf{I}\\\mathbf{-I} &\mathbf{0}\end{matrix} \right )
, 
\ \rm{where} \ \mathbf{I}=\left ( \begin{matrix}1 & 0\\0 &1\end{matrix} \right ).
\end{equation}

The parameter matrix of 2-D GFRFT is given by the following formula
\begin{equation}
    \mathbf{M}=\left ( \begin{matrix}\mathbf{A} & \mathbf{B}\\\mathbf{C} &\mathbf{D}\end{matrix} \right )=\left ( \begin{matrix}\mathbf{\Phi} & \mathbf{\Psi}\\\mathbf{-\Psi} &\mathbf{\Phi}\end{matrix} \right )
, 
\end{equation}
where $\mathbf{\Phi}=\left ( \begin{matrix}\cos \alpha & 0\\0 &\cos \alpha \end{matrix} \right )$ and $\mathbf{\Psi}=\left ( \begin{matrix}\sin \alpha & 0\\0 &\sin \alpha \end{matrix} \right )$.

The parameter matrix of 2-D graph CM is given by the following formula
\begin{equation}
    \mathbf{M}=\left ( \begin{matrix}\mathbf{A} & \mathbf{B}\\\mathbf{C} &\mathbf{D}\end{matrix} \right )=\left ( \begin{matrix}\mathbf{I} & \mathbf{0}\\\mathbf{G} &\mathbf{I}\end{matrix} \right )
, \ \rm{where} \ \mathbf{G}=\left ( \begin{matrix}\xi & 0\\0 &\xi\end{matrix} \right ).
\end{equation}

The parameter matrix of 2-D graph scaling transform is given by the following formula
\begin{equation}
    \mathbf{M}=\left ( \begin{matrix}\mathbf{A} & \mathbf{B}\\\mathbf{C} &\mathbf{D}\end{matrix} \right )=\left ( \begin{matrix}\mathbf{K} & \mathbf{0}\\\mathbf{0} &\mathbf{K^{-1}}\end{matrix} \right )
, \ \rm{where} \ \mathbf{K}=\left ( \begin{matrix}\sigma & 0\\0 &\sigma\end{matrix} \right ).
\end{equation}

Koc et al. \cite{Ko2010FastAA} proposed a fast algorithm which can be used to calculate 2-D GLCT. It is based on CDDHFs decomposition, and the 2-D parameter matrix decomposition is expressed as
\begin{equation}
    \begin{pmatrix}
		\mathbf{A}&\mathbf{B} \\
		\mathbf{C}&\mathbf{D}
	\end{pmatrix} = \begin{pmatrix}
		\mathbf{I}&\mathbf{0} \\
		\mathbf{G} &\mathbf{I}
	\end{pmatrix}
	\begin{pmatrix}
		\mathbf{K} &\mathbf{0} \\
		\mathbf{0}&\mathbf{K}^{-1}
	\end{pmatrix}
	\begin{pmatrix}
		\mathbf{\Phi } &\mathbf{\Psi }   \\
		-\mathbf{\Psi } &\mathbf{\Phi }
	\end{pmatrix}.
\end{equation}
The first matrix corresponds to 2-D graph CM with chirp rate $\xi $. The second matrix corresponds to 2-D graph scaling transform with scaling factor $\delta$. The third matrix corresponds to 2-D GFRFT with transform order $\alpha$. The relationship between parameter $\left ( \xi ,\delta ,\alpha  \right )$ and parameters $(a,b,c,d)$ is
\begin{equation}
\xi =\frac{ac+bd}{a^{2}+b^{2}},\delta =\sqrt{a^{2}+b^{2}},\alpha =\cos ^{-1}\left ( \frac{a}{\delta }  \right )= \sin  ^{-1}\left ( \frac{b}{\delta }  \right ).
\end{equation}
This decomposition is a special case of Iwasawa decomposition \cite{delaCruz2021OnTI} of symplectic matrix. Under this decomposition, 2-D GLCT can be realized by the combination of 2-D graph CM, 2-D graph scaling transform and 2-D GFRFT.

Definition 9. The 2-D CDDHFs-GLCT of signal $x:V_{1}\times V_{2}\longrightarrow \mathbb{R}$ on a Cartesian product graph $G_{1}\Box G_{2}$ is defined by decomposing of GSOs $\mathbf{Z}_{1},\mathbf{Z}_{2}$ as
\begin{equation}
\begin{split}
O_{\rm{2-D \ CDDHFs-GLCT}}^{(\mathbf{A,B;C ,D} )}\left ( \mathbf{x} \right )=& O_{\rm{2-D \ GCM}}^{\xi }O_{\rm{2-D \ GS}}^{\sigma  }O_{\rm{2-D \ GFRFT}}^{\alpha   }\left ( \mathbf{x} \right ) \\
=& \left( \mathbf{J}_{\mathbf{F}_{2} }\otimes \mathbf{J}_{\mathbf{F}_{1} } \right ) ^{\xi }\left ( \mathbf{V}_{{\mathbf{S}_{2} }}\otimes \mathbf{V}_{{\mathbf{S}_{1} }} \right ) \left ( \mathbf{J}_{\mathbf{S}_{2}}\otimes \mathbf{I}_{N_{1} } + \mathbf{I}_{N_{2} }\otimes \mathbf{J}_{\mathbf{S}_{1}} \right ) \\
&  \left ( {\mathbf{V}_{\mathbf{S}_{2} }^{-1} }\otimes {\mathbf{V}_{\mathbf{S}_{1} }^{-1} }\right ) \left ( \mathbf{V}_{\mathbf{F}_{2} }\otimes \mathbf{V}_{\mathbf{F}_{1} } \right ) \\
& \left ( \mathbf{J}_{\mathbf{F}_{2} }\otimes \mathbf{J}_{\mathbf{F}_{1} } \right ) ^{\alpha }\left ( \mathbf{V}_{\mathbf{F}_{2} }^{-1}\otimes \mathbf{V}_{\mathbf{F}_{1} }^{-1} \right ) \mathbf{x} \\
 = & \left( \mathbf{J}_{\mathbf{F}_{2} }\otimes \mathbf{J}_{\mathbf{F}_{1} } \right ) ^{\xi }\left ( \mathbf{V}_{{\mathbf{S}_{2} }}\otimes \mathbf{V}_{{\mathbf{S}_{1} }} \right )\left ( \mathbf{J}_{\mathbf{F}_{2} }\otimes \mathbf{J}_{\mathbf{F}_{1} } \right ) ^{\alpha } \\ 
 & \left ( \mathbf{V}_{\mathbf{F}_{2} }^{-1}\otimes \mathbf{V}_{\mathbf{F}_{1} }^{-1} \right ) \mathbf{x} ,
\end{split}
\end{equation}
where $\xi$ is the chirp rate, $\sigma$ is the scaling factor, and $\alpha $ is the fractional transform order. $O_{\rm{2-D \ CDDHFs-GLCT}}^{(\mathbf{A,B;C ,D} )}$ denotes the 2-D CDDHFs-GLCT operator.

\subsection{2-D CM-CC-CM-GLCT}
\subsubsection{when $\rm{det} \ \mathbf{B}\ne 0$}
Because 2-D discrete affine transforma introduce interpolation error, Pei et al. \cite{Pei2016FastDL} proposed a fast algorithm to calculate 2-D GLCT without scaling operation. It is based on CM-CC-CM decomposition, and the 2-D parameter matrix decomposition is expressed as
\begin{equation}
    \begin{pmatrix}
		\mathbf{A}&\mathbf{B} \\
		\mathbf{C}&\mathbf{D}
	\end{pmatrix} = \begin{pmatrix}
		\mathbf{I}&\mathbf{0} \\
		\mathbf{G}_{1} &\mathbf{I}
	\end{pmatrix}
	\begin{pmatrix}
		\mathbf{I} &\mathbf{B} \\
		\mathbf{0}&\mathbf{I}
	\end{pmatrix}
	\begin{pmatrix}
		\mathbf{I} &\mathbf{0}   \\
		\mathbf{G}_{3} &\mathbf{I}
	\end{pmatrix}.
\end{equation}

Obviously, $\mathbf{B}$ must be reversible. The first matrix corresponds to 2-D graph CM with chirp rate $\xi _{1}=\frac{d-1}{b}$, the second matrix corresponds to 2-D graph CC with parameter $b$, and the last matrix corresponds to 2D graph CM with chirp rate $\xi _{3}=\frac{a-1}{b}$. The 2D graph CC can be further decomposed, that is
\begin{equation}
    \begin{pmatrix}
		\mathbf{I}&\mathbf{B} \\
		\mathbf{0}&\mathbf{I}
	\end{pmatrix} = \begin{pmatrix}
		\mathbf{0}&\mathbf{-I} \\
		\mathbf{I} &\mathbf{0}
	\end{pmatrix}
	\begin{pmatrix}
		\mathbf{I} &\mathbf{0} \\
		\mathbf{G}_{2}&\mathbf{I}
	\end{pmatrix}
	\begin{pmatrix}
		\mathbf{0} &\mathbf{I}   \\
		-\mathbf{I} &\mathbf{0}
	\end{pmatrix},
\end{equation}
where $\xi _{2}=-b$. Then we can get the 2-D CM-CC-CM-GLCT which is completely composed of 2-D GFT, 2-D IGFT and 2-D graph CM.

Definition 9. The 2-D CM-CC-CM-GLCT $\left ( \rm{det}\ \mathbf{B}\ne 0 \right )$ of signal $x:V_{1}\times V_{2}\longrightarrow \mathbb{R}$ on a Cartesian product graph $G_{1}\Box G_{2}$ is defined by decomposing of GSOs $\mathbf{Z}_{1},\mathbf{Z}_{2}$ as
\begin{equation}
\begin{split}
O_{\rm{2-D \ CM-CC-CM-GLCT}}^{(\mathbf{A,B;C ,D} )}\left ( \mathbf{x} \right )=& O_{\rm{2-D \ GCM}}^{\xi_{1}  }O_{\rm{2-D \ IGFT}}O_{\rm{2-D \ GCM}}^{\xi_{2}  } \\ 
& O_{\rm{2-D \ GFT}}O_{\rm{2-D \ GCM}}^{\xi_{3}  } \left ( \mathbf{x} \right ) \\
=& \left( \mathbf{J}_{\mathbf{F}_{2} }\otimes \mathbf{J}_{\mathbf{F}_{1} } \right ) ^{\xi _{1} }\left ( \mathbf{F}_{{\mathbf{G}_{2} }}^{-1}\otimes \mathbf{F}_{{\mathbf{G}_{1} }}^{-1}  \right ) \left( \mathbf{J}_{\mathbf{F}_{2} }\otimes \mathbf{J}_{\mathbf{F}_{1} } \right ) ^{\xi _{2} } \\ 
& \left ( \mathbf{F}_{{\mathbf{G}_{2} }}\otimes \mathbf{F}_{{\mathbf{G}_{1} }}  \right )\left( \mathbf{J}_{\mathbf{F}_{2} }\otimes \mathbf{J}_{\mathbf{F}_{1} } \right ) ^{\xi _{3} } \mathbf{x} ,
\end{split}
\end{equation}
where $O_{\rm{2-D \ CM-CC-CM-GLCT}}^{(\mathbf{A,B;C ,D} )}$ denotes the 2-D CM-CC-CM-GLCT operator. The relationship between parameter $\left (\xi _{1} , \xi _{2} , \xi _{3 } \right )$ and parameters $(a,b,c,d)$ is
\begin{equation}
\xi _{1}=\frac{d-1}{b},\xi _{2}=-b,\xi _{3}=\frac{a-1}{b}.
\end{equation}
\subsubsection{when $\rm{det} \ \mathbf{B}= 0$}
By observing the parameters, we can find that the definition in formula (26) is invalid when $\mathbf{B}=0$, that is $b=0$. In the case of $b=0$, since the parameter satisfies $ad-bc=1$, there are $a\ne 0$ and $d\ne 0$. Then the following two kinds of decomposition can be considered \cite{Pei2016FastDL}
\begin{equation}
\begin{pmatrix}
	\mathbf{A}&\mathbf{0} \\
	\mathbf{C} &\mathbf{D}
\end{pmatrix}=
\begin{pmatrix}
	\mathbf{0}&\mathbf{I} \\
	\mathbf{-I} &\mathbf{0}
\end{pmatrix}
\begin{pmatrix}
	\mathbf{I}&\mathbf{0} \\
	\mathbf{G}_{4} &\mathbf{I} 
\end{pmatrix}
\begin{pmatrix}
	\mathbf{0}&\mathbf{-I} \\
	\mathbf{I} &\mathbf{0}
\end{pmatrix}
\begin{pmatrix}
	\mathbf{I}&\mathbf{0} \\
	\mathbf{G}_{5} &\mathbf{I}
\end{pmatrix}
\begin{pmatrix}
	\mathbf{0}&\mathbf{I} \\
	-\mathbf{I} &\mathbf{0}
\end{pmatrix}
\begin{pmatrix}
	\mathbf{I}&\mathbf{0} \\
	\mathbf{G}_{6} &\mathbf{I}
\end{pmatrix},
\end{equation}
and
\begin{equation}
\begin{pmatrix}
	\mathbf{A}&\mathbf{0} \\
	\mathbf{C} &\mathbf{D}
\end{pmatrix}=
\begin{pmatrix}
	\mathbf{I}&\mathbf{0} \\
	\mathbf{G}_{7} &\mathbf{I}
\end{pmatrix}
\begin{pmatrix}
	\mathbf{0}&-\mathbf{I} \\
	\mathbf{I} &\mathbf{0} 
\end{pmatrix}
\begin{pmatrix}
	\mathbf{I}&\mathbf{0} \\
	\mathbf{G}_{8} &\mathbf{I}
\end{pmatrix}
\begin{pmatrix}
	\mathbf{0}&\mathbf{I} \\
	-\mathbf{I} &\mathbf{0}
\end{pmatrix}
\begin{pmatrix}
	\mathbf{I}&\mathbf{0} \\
	\mathbf{G}_{9} &\mathbf{I}
\end{pmatrix}
\begin{pmatrix}
	\mathbf{0}&-\mathbf{I} \\
	\mathbf{I} &\mathbf{0}
\end{pmatrix}.
\end{equation}

Definition 10. The 2-D CM-CC-CM-GLCT $\left ( \rm{det}\ \mathbf{B}= 0 \right )$ of signal $x:V_{1}\times V_{2}\longrightarrow \mathbb{R}$ on a Cartesian product graph $G_{1}\Box G_{2}$ is defined by decomposing of GSOs $\mathbf{Z}_{1},\mathbf{Z}_{2}$ as
\begin{equation}
\begin{split}
O_{\rm{2-D \ CM-CC-CM-GLCT} }^{\left ( \mathbf{A,0,C,D} \right )}\left ( \mathbf{x} \right )  = & \sqrt{-\rm{j}} \left ( \mathbf{F}_{{\mathbf{G}_{2} }}\otimes \mathbf{F}_{{\mathbf{G}_{1} }}  \right )\left( \mathbf{J}_{\mathbf{F}_{2} }\otimes \mathbf{J}_{\mathbf{F}_{1} } \right ) ^{\xi _{4} }\left ( \mathbf{F}_{{\mathbf{G}_{2} }}^{-1}\otimes \mathbf{F}_{{\mathbf{G}_{1} }}^{-1}  \right ) \\
& \left( \mathbf{J}_{\mathbf{F}_{2} }\otimes \mathbf{J}_{\mathbf{F}_{1} } \right ) ^{\xi _{5} } \left ( \mathbf{F}_{{\mathbf{G}_{2} }}\otimes \mathbf{F}_{{\mathbf{G}_{1} }}  \right )\left( \mathbf{J}_{\mathbf{F}_{2} }\otimes \mathbf{J}_{\mathbf{F}_{1} } \right ) ^{\xi _{6} } \mathbf{x},    
\end{split}
\end{equation}
\begin{equation}
\begin{split}
O_{\rm{2-D \ CM-CC-CM-GLCT} }^{\left ( \mathbf{A,0,C,D} \right )}\left ( \mathbf{x} \right )  = & \sqrt{\rm{j}} \left( \mathbf{J}_{\mathbf{F}_{2} }\otimes \mathbf{J}_{\mathbf{F}_{1} } \right ) ^{\xi _{7} }\left ( \mathbf{F}_{{\mathbf{G}_{2} }}^{-1}\otimes \mathbf{F}_{{\mathbf{G}_{1} }}^{-1}  \right )\left( \mathbf{J}_{\mathbf{F}_{2} }\otimes \mathbf{J}_{\mathbf{F}_{1} } \right ) ^{\xi _{8} } \\ 
& \left ( \mathbf{F}_{{\mathbf{G}_{2} }}\otimes \mathbf{F}_{{\mathbf{G}_{1} }}  \right )\left( \mathbf{J}_{\mathbf{F}_{2} }\otimes \mathbf{J}_{\mathbf{F}_{1} } \right ) ^{\xi _{9} }\left ( \mathbf{F}_{{\mathbf{G}_{2} }}^{-1}\otimes \mathbf{F}_{{\mathbf{G}_{1} }}^{-1}  \right ) \mathbf{x},    
\end{split}
\end{equation}
where $\sqrt{\rm{j}}$ is the phase difference between FT and LCT, and $\sqrt{-\rm{j}}$ is the phase difference between IFT and LCT.
The relationship between parameter $\left (\xi  _{4} , \xi  _{5} , \xi  _{6 }  \right )$ and parameters $(a,b,c,d)$ is
\begin{equation}
\xi  _{4}=\frac{1}{d},\xi  _{5}=d,\xi  _{6}=\frac{c+1}{d}.
\end{equation}
The relationship between parameter $\left (\xi  _{7} , \xi  _{8} , \xi  _{9 }  \right )$ and parameters $(a,b,c,d)$ is
\begin{equation}
\xi  _{7}=\frac{c-1}{a},\xi  _{8}=-a,\xi  _{9}=-\frac{1}{a}.
\end{equation}

\section{ Multi-dimensional graph linear canonical transform }
\subsection{M-D CDDHFs-GLCT}
Definition 11. The M-D CDDHFs-GLCT of a M-D signal $\mathbf{x}$ on a Cartesian product graph $G_{1}\Box \dots \Box G_{m}$ is defined by decomposing of GSOs $\mathbf{Z}_{1},\dots,\mathbf{Z}_{m}$ as
\begin{equation}
\begin{split}
O_{\rm{M-D \ CDDHFs-GLCT}}^{(\mathbf{A,B;C ,D} )}\left ( \mathbf{x} \right )=& O_{\rm{M-D \ GCM}}^{\xi }O_{\rm{M-D \ GS}}^{\sigma  }O_{\rm{M-D \ GFRFT}}^{\alpha   }\left ( \mathbf{x} \right ) \\
 = & \left( \mathbf{J}_{\mathbf{F}_{m} }\otimes \dots \otimes \mathbf{J}_{\mathbf{F}_{1} } \right ) ^{\xi }\left ( \mathbf{V}_{{\mathbf{S}_{m} }}\otimes \dots \otimes \mathbf{V}_{{\mathbf{S}_{1} }} \right )\\
 & \left ( \mathbf{J}_{\mathbf{F}_{m} }\otimes \dots \otimes \mathbf{J}_{\mathbf{F}_{1} } \right ) ^{\alpha }\left ( \mathbf{V}_{\mathbf{F}_{m} }^{-1}\otimes \dots \otimes \mathbf{V}_{\mathbf{F}_{1} }^{-1} \right ) \mathbf{x}, 
\end{split}
\end{equation}
where $\xi$ is the chirp rate, $\sigma$ is the scaling factor, and $\alpha $ is the fractional transform order. $O_{\rm{M-D \ CDDHFs-GLCT}}^{(\mathbf{A,B;C ,D} )}$ denotes the M-D CDDHFs-GLCT operator. The relationship between parameter $\left ( \xi ,\delta ,\alpha  \right )$ and parameters $(a,b,c,d)$ is
\begin{equation}
\xi =\frac{ac+bd}{a^{2}+b^{2}},\delta =\sqrt{a^{2}+b^{2}},\alpha =\cos ^{-1}\left ( \frac{a}{\delta }  \right )= \sin  ^{-1}\left ( \frac{b}{\delta }  \right ).
\end{equation}

\subsection{M-D CM-CC-CM-GLCT}
\subsubsection{when $\rm{det} \ \mathbf{B}\ne 0$}
Definition 12. The M-D CM-CC-CM-GLCT $\left ( \rm{det} \ \mathbf{B}\ne 0 \right )$ of a M-D signal $\mathbf{x}$ on a Cartesian product graph $G_{1}\Box \dots \Box G_{m}$ is defined by decomposing of GSOs $\mathbf{Z}_{1},\dots,\mathbf{Z}_{m}$ as
\begin{equation}
\begin{split}
O_{\rm{M-D \ CM-CC-CM-GLCT}}^{(\mathbf{A,B;C ,D} )}\left ( \mathbf{x} \right )=& O_{\rm{M-D \ GCM}}^{\xi_{1}  }O_{\rm{M-D \ IGFT}}O_{\rm{M-D \ GCM}}^{\xi_{2}  } \\ 
& O_{\rm{M-D \ GFT}}O_{\rm{M-D \ GCM}}^{\xi_{3}  }\left ( \mathbf{x} \right ) \\
=& \left( \mathbf{J}_{\mathbf{F}_{m} }\otimes \dots \otimes \mathbf{J}_{\mathbf{F}_{1} } \right ) ^{\xi _{1} }\left ( \mathbf{F}_{{\mathbf{G}_{m} }}^{-1}\otimes \dots \otimes \mathbf{F}_{{\mathbf{G}_{1} }}^{-1}  \right ) \\
&\left( \mathbf{J}_{\mathbf{F}_{m} }\otimes \dots \otimes \mathbf{J}_{\mathbf{F}_{1} } \right ) ^{\xi _{2} } \left ( \mathbf{F}_{{\mathbf{G}_{m} }}\otimes \dots \otimes \mathbf{F}_{{\mathbf{G}_{1} }}  \right ) \\ 
& \left( \mathbf{J}_{\mathbf{F}_{m} }\otimes \dots \otimes \mathbf{J}_{\mathbf{F}_{1} } \right ) ^{\xi _{3} } \mathbf{x} ,
\end{split}
\end{equation}
where $O_{\rm{M-D \ CM-CC-CM-GLCT}}^{(\mathbf{A,B;C ,D} )}$ denotes the M-D CM-CC-CM-GLCT operator. The relationship between parameter $\left (\xi _{1} , \xi _{2} , \xi _{3 } \right )$ and parameters $(a,b,c,d)$ is
\begin{equation}
\xi _{1}=\frac{d-1}{b},\xi _{2}=-b,\xi _{3}=\frac{a-1}{b}.
\end{equation}

\subsubsection{when $\rm{det} \ \mathbf{B}= 0$}
Definition 13. The M-D CM-CC-CM-GLCT $\left ( \rm{det} \ \mathbf{B}= 0 \right ) $ of a M-D signal $\mathbf{x}$ on a Cartesian product graph $G_{1}\Box \dots \Box G_{m}$ is defined by decomposing of GSOs $\mathbf{Z}_{1},\dots,\mathbf{Z}_{m}$ as
\begin{equation}
\begin{split}
O_{\rm{M-D \ CM-CC-CM-GLCT} }^{\left ( \mathbf{A,0,C,D} \right )}\left ( \mathbf{x} \right )  = & \sqrt{-\rm{j}} \left ( \mathbf{F}_{{\mathbf{G}_{m} }}\otimes \dots \otimes 
 \mathbf{F}_{{\mathbf{G}_{1} }}  \right )\left( \mathbf{J}_{\mathbf{F}_{m} }\otimes \dots \otimes \mathbf{J}_{\mathbf{F}_{1} } \right ) ^{\xi _{4} } \\ 
 & \left ( \mathbf{F}_{{\mathbf{G}_{m} }}^{-1}\otimes \dots \otimes  \mathbf{F}_{{\mathbf{G}_{1} }}^{-1}  \right )  \left( \mathbf{J}_{\mathbf{F}_{m} }\otimes \dots \otimes  \mathbf{J}_{\mathbf{F}_{1} } \right ) ^{\xi _{5} } \\
 & \left ( \mathbf{F}_{{\mathbf{G}_{m} }}\otimes \dots \otimes \mathbf{F}_{{\mathbf{G}_{1} }}  \right )\left( \mathbf{J}_{\mathbf{F}_{m} }\otimes \dots \otimes \mathbf{J}_{\mathbf{F}_{1} } \right ) ^{\xi _{6} } \mathbf{x},    
\end{split}
\end{equation}
\begin{equation}
\begin{split}
O_{\rm{M-D \ CM-CC-CM-GLCT} }^{\left ( \mathbf{A,0,C,D} \right )}\left ( \mathbf{x} \right )  = & \sqrt{\rm{j}} \left( \mathbf{J}_{\mathbf{F}_{m} }\otimes \dots \otimes 
 \mathbf{J}_{\mathbf{F}_{1} } \right ) ^{\xi _{7} }\left ( \mathbf{F}_{{\mathbf{G}_{m} }}^{-1}\otimes \dots \otimes \mathbf{F}_{{\mathbf{G}_{1} }}^{-1}  \right ) \\
 & \left( \mathbf{J}_{\mathbf{F}_{m} }\otimes \dots \otimes \mathbf{J}_{\mathbf{F}_{1} } \right ) ^{\xi _{8} } \left ( \mathbf{F}_{{\mathbf{G}_{m} }}\otimes \dots \otimes \mathbf{F}_{{\mathbf{G}_{1} }}  \right ) \\
 & \left( \mathbf{J}_{\mathbf{F}_{m} }\otimes \dots \otimes \mathbf{J}_{\mathbf{F}_{1} } \right ) ^{\xi _{9} }\left ( \mathbf{F}_{{\mathbf{G}_{m} }}^{-1}\otimes \dots \otimes  \mathbf{F}_{{\mathbf{G}_{1} }}^{-1}  \right ) \mathbf{x},    
\end{split}
\end{equation}
where $\sqrt{\rm{j}}$ is the phase difference between FT and LCT, and $\sqrt{-\rm{j}}$ is the phase difference between IFT and LCT.
The relationship between parameter $\left (\xi  _{4} , \xi  _{5} , \xi  _{6 }  \right )$ and parameters $(a,b,c,d)$ is
\begin{equation}
\xi  _{4}=\frac{1}{d},\xi  _{5}=d,\xi  _{6}=\frac{c+1}{d}.
\end{equation}
The relationship between parameter $\left (\xi  _{7} , \xi  _{8} , \xi  _{9 }  \right )$ and parameters $(a,b,c,d)$ is
\begin{equation}
\xi  _{7}=\frac{c-1}{a},\xi  _{8}=-a,\xi  _{9}=-\frac{1}{a}.
\end{equation}

\section{Comparison between M-D CDDHFS-GLCT and M-D CM-CC-CM-GLCT}
In this section, the M-D CDDHFS-GLCT is compared to the M-D CM-CC-CM-GLCT in terms of computational complexity, additivity, and reversibility. Both be irrelevant to sampling periods and without oversampling operation. In order to facilitate the comparative discussion, we take the 2-D case as an example.
\subsection{Computational complexity}
Recall the matrix form of the 2-D CDDHFs-GLCT showm in (23), i.e., $O_{\rm{2-D \ CDDHFs-GLCT}}^{(\mathbf{A,B;C ,D} )} = \\ \left( \mathbf{J}_{\mathbf{F}_{2} }\otimes \mathbf{J}_{\mathbf{F}_{1} } \right ) ^{\xi }\left ( \mathbf{V}_{{\mathbf{S}_{2} }}\otimes \mathbf{V}_{{\mathbf{S}_{1} }} \right )\left ( \mathbf{J}_{\mathbf{F}_{2} }\otimes \mathbf{J}_{\mathbf{F}_{1} } \right ) ^{\alpha } \left ( \mathbf{V}_{\mathbf{F}_{2} }^{-1}\otimes \mathbf{V}_{\mathbf{F}_{1} }^{-1} \right ) = \left( \mathbf{J}_{\mathbf{F}_{2} }^{\xi }\otimes \mathbf{J}_{\mathbf{F}_{1} }^{\xi } \right ) \left ( \mathbf{V}_{{\mathbf{S}_{2} }}\otimes \mathbf{V}_{{\mathbf{S}_{1} }} \right )\left ( \mathbf{J}_{\mathbf{F}_{2} }^{\alpha }\otimes \mathbf{J}_{\mathbf{F}_{1} }^{\alpha } \right ) \\ \left ( \mathbf{V}_{\mathbf{F}_{2} }^{-1}\otimes \mathbf{V}_{\mathbf{F}_{1} }^{-1} \right )$. Each of the real matrix $\mathbf{V}_{{\mathbf{S}_{1} }}$, $\mathbf{V}_{\mathbf{F}_{1} }^{-1}$ generates $2N_{1}^{2} $ real multiplications, and $\mathbf{V}_{{\mathbf{S}_{2} }}$, $\mathbf{V}_{\mathbf{F}_{2} }^{-1}$ generates $2N_{2}^{2} $ real multiplications. Each of complex diagonal matrix $\mathbf{J}_{\mathbf{F}_{1} }^{\xi }$, $ \mathbf{J}_{\mathbf{F}_{1} }^{\alpha }$ generates $N_{1}$ complex multiplications, and $\mathbf{J}_{\mathbf{F}_{2} }^{\xi }$, $ \mathbf{J}_{\mathbf{F}_{2} }^{\alpha }$ generates $N_{2}$ complex multiplications. Therefore, the computation totally includes $4\left ( N_{1}^{2} + N_{2}^{2} \right )+8\left ( N_{1} + N_{2} \right ) $ real multiplications. Regarding 2-D CM-CC-CM-GLCT, firstly consider the $\rm{det} \ \mathbf{B} \ne 0$ case. The 2-D CM-CC-CM-GLCT shown in (26), that is, $O_{\rm{2-D \ CM-CC-CM-GLCT}}^{(\mathbf{A,B;C ,D} )}= \left( \mathbf{J}_{\mathbf{F}_{2} }\otimes \mathbf{J}_{\mathbf{F}_{1} } \right ) ^{\xi _{1} }\left ( \mathbf{F}_{{\mathbf{G}_{2} }}^{-1}\otimes \mathbf{F}_{{\mathbf{G}_{1} }}^{-1}  \right ) \left( \mathbf{J}_{\mathbf{F}_{2} }\otimes \mathbf{J}_{\mathbf{F}_{1} } \right ) ^{\xi _{2} } \left ( \mathbf{F}_{{\mathbf{G}_{2} }}\otimes \mathbf{F}_{{\mathbf{G}_{1} }}  \right ) \\ \left( \mathbf{J}_{\mathbf{F}_{2} }\otimes \mathbf{J}_{\mathbf{F}_{1} } \right ) ^{\xi _{3} }$. There are three 2-D graph CMs which require $3\left ( N_{1} + N_{2} \right )$ complex multiplications. The 2-D GFT and 2-D IGFT can be calculated by FFT, which requires  $\left ( N_{1} /2 \right )\log_{2}{N_{1} } + \left ( N_{2} /2 \right )\log_{2}{N_{2} }$ complex multiplications respectively. Therefore, the computational complexity of 2-D CM-CC-CM-GLCT for $\rm{det} \ \mathbf{B} \ne 0$ is $12\left ( N_{1} + N_{2} \right ) +4\left ( N_{1}\log_{2}{N_{1} } + N_{2}\log_{2}{N_{2} } \right )$ real multiplications. The 2-D CM-CC-CM-GLCT for $\rm{det} \ \mathbf{B} = 0$ shown in (30) and (31), which contains one more 2-D GFT (or 2-D IGFT) than the $\rm{det} \ \mathbf{B} \ne 0$ case. Therefore, its complexity increases to $12\left ( N_{1} + N_{2} \right ) +6\left ( N_{1}\log_{2}{N_{1} } + N_{2}\log_{2}{N_{2} } \right )$ real multiplications. Finally, we comprehensively compare the computational complexity in Table 1.

\begin{table}[]
	\caption{Complexity of 2-D CDDHFs-GLCT and 2-D CM-CC-CM-GLCT}
	\centering
	\begin{threeparttable}
        \resizebox{\textwidth}{!}{
	\begin{tabular}{|l|l|l|}
		\hline
		& Matrix form                                                                                                                                                                                                                                                                                      & Complexity                         \\ \hline
		2-D CDDHFs-GLCT                            & $\left( \mathbf{J}_{\mathbf{F}_{2} }\otimes \mathbf{J}_{\mathbf{F}_{1} } \right ) ^{\xi }\left ( \mathbf{V}_{{\mathbf{S}_{2} }}\otimes \mathbf{V}_{{\mathbf{S}_{1} }} \right )\left ( \mathbf{J}_{\mathbf{F}_{2} }\otimes \mathbf{J}_{\mathbf{F}_{1} } \right ) ^{\alpha }\left ( \mathbf{V}_{\mathbf{F}_{2} }^{-1}\otimes \mathbf{V}_{\mathbf{F}_{1} }^{-1} \right )$                                                                                                                                                                                                                                                    & $ 4\left ( N_{1}^{2} + N_{2}^{2} \right ) +8\left ( N_{1} + N_{2} \right )$                        \\ \hline
		 2-D CM-CC-CM-GLCT $(\rm{det} \ \mathbf{B}\ne 0)$                & $\left( \mathbf{J}_{\mathbf{F}_{2} }\otimes \mathbf{J}_{\mathbf{F}_{1} } \right ) ^{\xi _{1} }\left ( \mathbf{F}_{{\mathbf{G}_{2} }}^{-1}\otimes \mathbf{F}_{{\mathbf{G}_{1} }}^{-1}  \right ) \left( \mathbf{J}_{\mathbf{F}_{2} }\otimes \mathbf{J}_{\mathbf{F}_{1} } \right ) ^{\xi _{2} } \left ( \mathbf{F}_{{\mathbf{G}_{2} }}\otimes \mathbf{F}_{{\mathbf{G}_{1} }}  \right )  \left( \mathbf{J}_{\mathbf{F}_{2} }\otimes \mathbf{J}_{\mathbf{F}_{1} } \right ) ^{\xi _{3} }$                                                                                                                                                                                                                                     & $12\left ( N_{1} + N_{2} \right ) +4\left ( N_{1}\log_{2}{N_{1} } + N_{2}\log_{2}{N_{2} } \right )$                 \\ \hline
		2-D CM-CC-CM-GLCT $(\rm{det} \ \mathbf{B}=0)$  & \multirow{2}{*}{\begin{tabular}[c]{@{}l@{}}$\sqrt{-j} \left ( \mathbf{F}_{{\mathbf{G}_{2} }}\otimes \mathbf{F}_{{\mathbf{G}_{1} }}  \right )\left( \mathbf{J}_{\mathbf{F}_{2} }\otimes \mathbf{J}_{\mathbf{F}_{1} } \right ) ^{\xi _{4} }\left ( \mathbf{F}_{{\mathbf{G}_{2} }}^{-1}\otimes \mathbf{F}_{{\mathbf{G}_{1} }}^{-1}  \right ) \left( \mathbf{J}_{\mathbf{F}_{2} }\otimes \mathbf{J}_{\mathbf{F}_{1} } \right ) ^{\xi _{5} } \left ( \mathbf{F}_{{\mathbf{G}_{2} }}\otimes \mathbf{F}_{{\mathbf{G}_{1} }}  \right )\left( \mathbf{J}_{\mathbf{F}_{2} }\otimes \mathbf{J}_{\mathbf{F}_{1} } \right ) ^{\xi _{6} }$\\ 
  $\sqrt{j} \left( \mathbf{J}_{\mathbf{F}_{2} }\otimes \mathbf{J}_{\mathbf{F}_{1} } \right ) ^{\xi _{7} }\left ( \mathbf{F}_{{\mathbf{G}_{2} }}^{-1}\otimes \mathbf{F}_{{\mathbf{G}_{1} }}^{-1}  \right )\left( \mathbf{J}_{\mathbf{F}_{2} }\otimes \mathbf{J}_{\mathbf{F}_{1} } \right ) ^{\xi _{8} }\left ( \mathbf{F}_{{\mathbf{G}_{2} }}\otimes \mathbf{F}_{{\mathbf{G}_{1} }}  \right )\left( \mathbf{J}_{\mathbf{F}_{2} }\otimes \mathbf{J}_{\mathbf{F}_{1} } \right ) ^{\xi _{9} }\left ( \mathbf{F}_{{\mathbf{G}_{2} }}^{-1}\otimes \mathbf{F}_{{\mathbf{G}_{1} }}^{-1}  \right )$\end{tabular}} & \multirow{2}{*}{ $12\left ( N_{1} + N_{2} \right ) +6\left ( N_{1}\log_{2}{N_{1} } + N_{2}\log_{2}{N_{2} } \right )$} \\
		&                                                                                                                                                                                                                                                                                                  &                                    \\ \hline
	\end{tabular}
}

	\begin{tablenotes}
		\footnotesize
		\item[*]  The complexity of matrix  eigendecomposition for $\mathbf{V}_{{\mathbf{S}_{1} }}$, $\mathbf{V}_{{\mathbf{S}_{2} }}$, $\mathbf{V}_{\mathbf{F}_{1} }^{-1}$ and $\mathbf{V}_{\mathbf{F}_{2} }^{-1}$ are not included.
	\end{tablenotes}

\end{threeparttable}

\end{table}

\subsection{Additivity property}
The additivity property of the 2-D GLCT is expressed as matrix multiplication, and the additivity property of the 2-D GLCT represents
\begin{equation}
	O_{\rm{2-D \ GLCT}}^{\mathbf{M}_{1} }O_{\rm{2-D \ GLCT}}^{\mathbf{M}_{2} }=O_{\rm{2-D \ GLCT}}^{\mathbf{M}_{1}\times \mathbf{M}_{2}  },
\end{equation}
where $\mathbf{M}_{1}=\begin{pmatrix}
	\mathbf{A}_{1} &\mathbf{B}_{1}  \\
	\mathbf{C}_{1}&\mathbf{D}_{1}
\end{pmatrix}  $, $\mathbf{M}_{2}=\begin{pmatrix}
\mathbf{A}_{2} &\mathbf{B}_{2}  \\
	\mathbf{C}_{2}&\mathbf{D}_{2}
\end{pmatrix}  $.

In the following, we compare the NMSE of the additivity property for 2-D CDDHFs-GLCT and 2-D CM-CC-CM-GLCT \cite{Pei2016FastDL}:
\begin{equation}
	{\rm NMSE}=\frac{\sum_{n=0}^{N_{1}N_{2}-1}\left | O_{\rm{2-D \ GLCT}}^{\mathbf{M}_{1}\times \mathbf{M}_{2} }\left ( x\left [ n \right ]  \right )  - O_{\rm{2-D \ GLCT}}^{\mathbf{ M}_{1}}O_{\rm{2-D \ GLCT}}^{\mathbf{M}_{2}}\left ( x\left [ n \right ]  \right )    \right |^{2} }{\sum_{n=0}^{N_{1}N_{2}-1}\left | O_{\rm{2-D \ GLCT}}^{\mathbf{M}_{1}\times \mathbf{M}_{2} }\left ( x\left [ n \right ]  \right )   \right |^{2}   } .
\end{equation}
The simulation is carried out on 2-D graph signals. Four kinds of 2-D graph signals are shown in Figure 1 and described as follows: \\
(1) $\mathbf{x}_{1}$: $\mathbf{x}_{1}^{\left ( 1 \right ) } =\left \{ x_{i_{1} }^{\left ( 1 \right )}  \right \}$ is a bipolar rectangular signal on the ring graph $G_{1}^{\left ( 1 \right ) } $ with 14 vertices and $\mathbf{x}_{1}^{\left ( 2 \right ) } =\left \{ x_{i_{2} }^{\left ( 2 \right )}  \right \}$ is a bipolar rectangular signal on the path graph $G_{1}^{\left ( 2 \right ) }$ with 8 vertices.\\
(2) $\mathbf{x}_{2}$: $\mathbf{x}_{2}^{\left ( 1 \right ) } =\left \{ x_{i_{1} }^{\left ( 1 \right )}  \right \}$ is a bipolar rectangular signal on the ring graph $G_{2}^{\left ( 1 \right ) } $ with 18 vertices and $\mathbf{x}_{2}^{\left (2 \right ) } =\left \{ x_{i_{2} }^{\left (2 \right )}  \right \}$ is a bipolar rectangular signal on the low stretch tree $G_{2}^{\left ( 2 \right ) }$ with 16 vertices. \\
(3) $\mathbf{x}_{3}$:  $\mathbf{x}_{3}^{\left ( 1 \right ) } =\left \{ x_{i_{1} }^{\left ( 1 \right )}  \right \}$ is a bipolar rectangular signal on the fully connected graph $G_{3}^{\left ( 1 \right ) } $ with 14 vertices and $\mathbf{x}_{3}^{\left ( 2 \right ) } =\left \{ x_{i_{2} }^{\left ( 2 \right )}  \right \}$ is a bipolar rectangular signal on the comet graph $G_{3}^{\left ( 2 \right ) }$ with 6 vertices. \\
(4) $\mathbf{x}_{4}$: $\mathbf{x}_{4}^{\left ( 1 \right ) } =\left \{ x_{i_{1} }^{\left ( 1 \right )}  \right \}$, where $\mathbf{x}_{41}$ is a bipolar rectangular signal on the fully connected graph $G_{4}^{\left ( 1 \right ) } $ with 8 vertices and $\mathbf{x}_{4}^{\left ( 2 \right ) } =\left \{ x_{i_{2} }^{\left (2 \right )}  \right \}$ is a bipolar rectangular signal on the low stretch tree $G_{4}^{\left ( 2 \right ) }$ with 16 vertices. 

The parameters in $\mathbf{M}_{1}$ and $\mathbf{M}_{2}$ are random numbers uniformly distributed on the interval $\left [ -2,2 \right ]$, which are used for 50 simulation runs. The additivity NMSE results in (43) is obtained. The NMSEs sorted in ascending order using $\mathbf{x}_{1}-\mathbf{x}_{4}$ as input signals are shown in Figure 2 respectively. In addition, the numerical verification of the average value of NMSEs for these graph signals for running 1000 times is shown in Table 2. These examples reveal that 2-D CM-CC-CM-GLCT has similar performance to the 2-D CDDHFs-GLCT in the additivity property.

\begin{figure}
\subfigure[]{
	\includegraphics[scale=0.4]{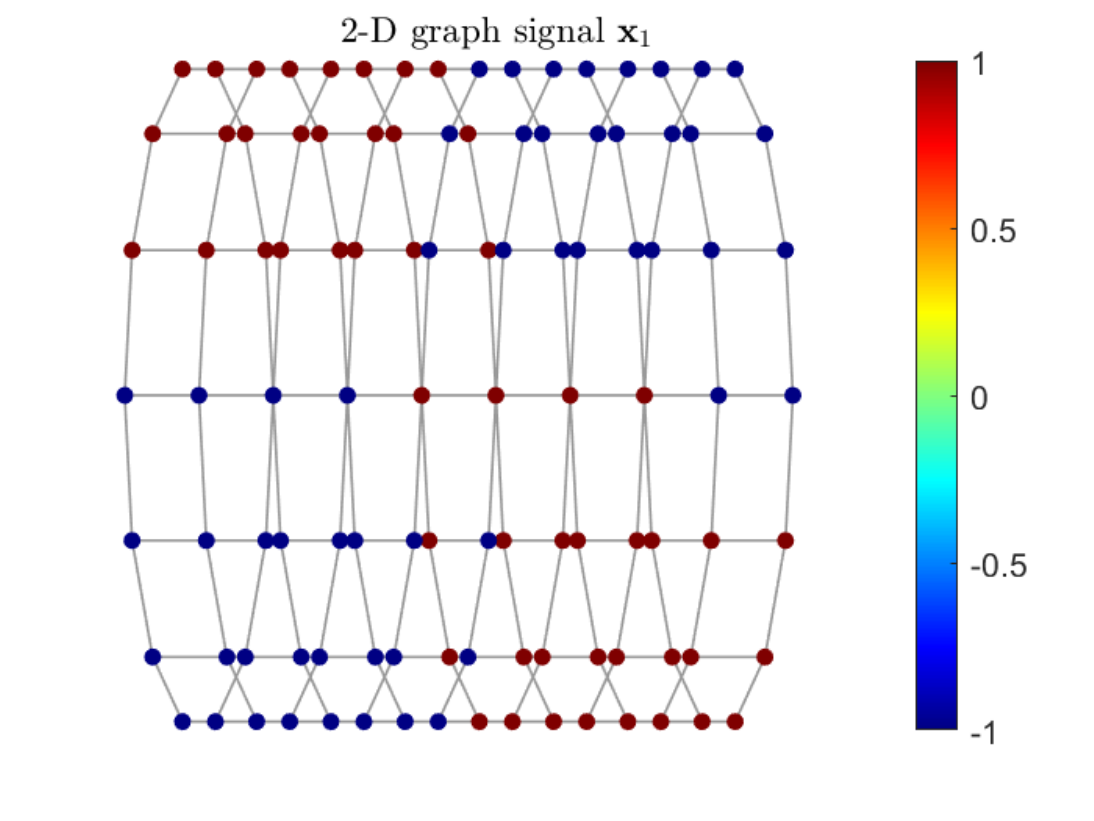} \label{1}
}
\quad
\subfigure[]{
	\includegraphics[scale=0.4]{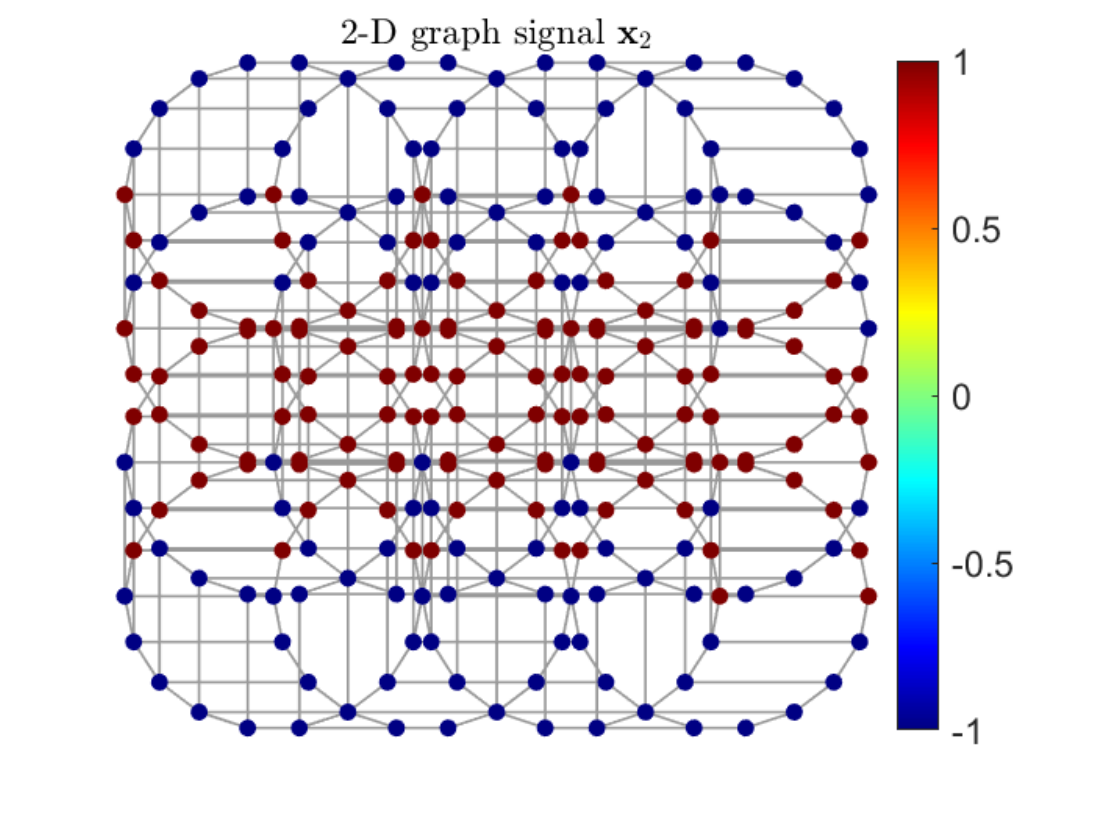} \label{2} 
}
\quad
\subfigure[]{
	\includegraphics[scale=0.4]{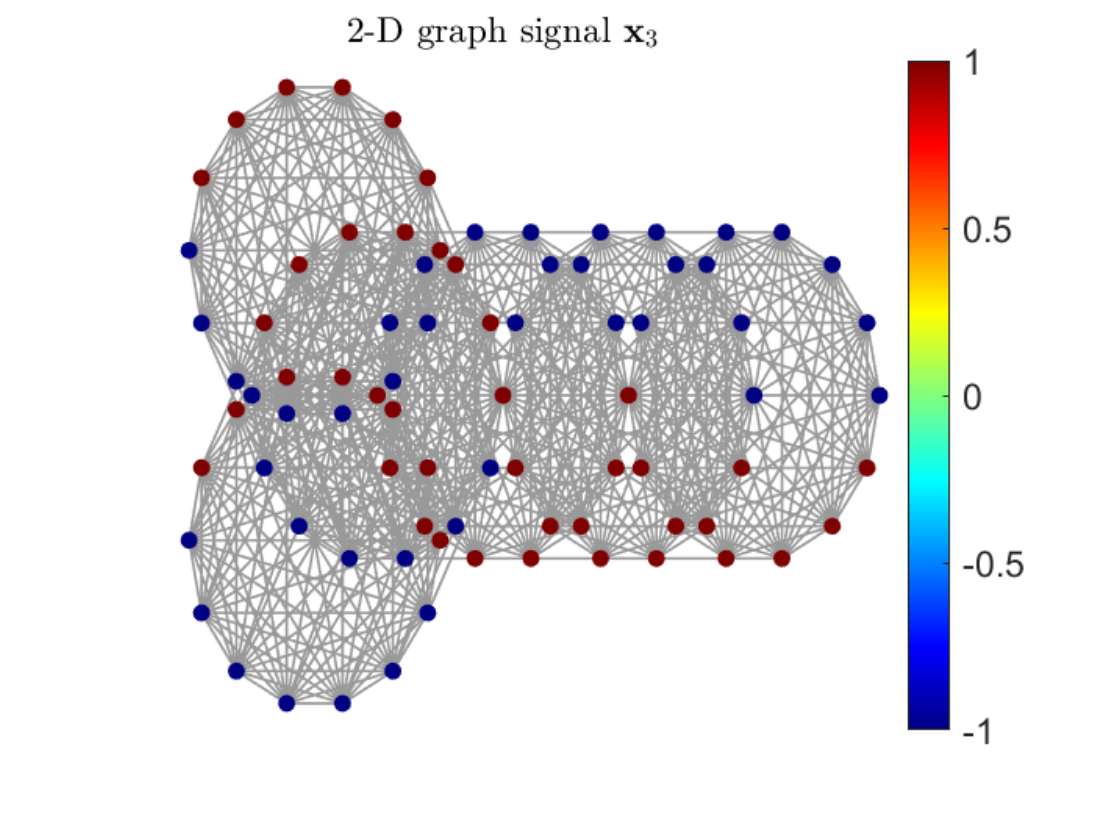} \label{3} 
}
\quad
\subfigure[]{
	\includegraphics[scale=0.4]{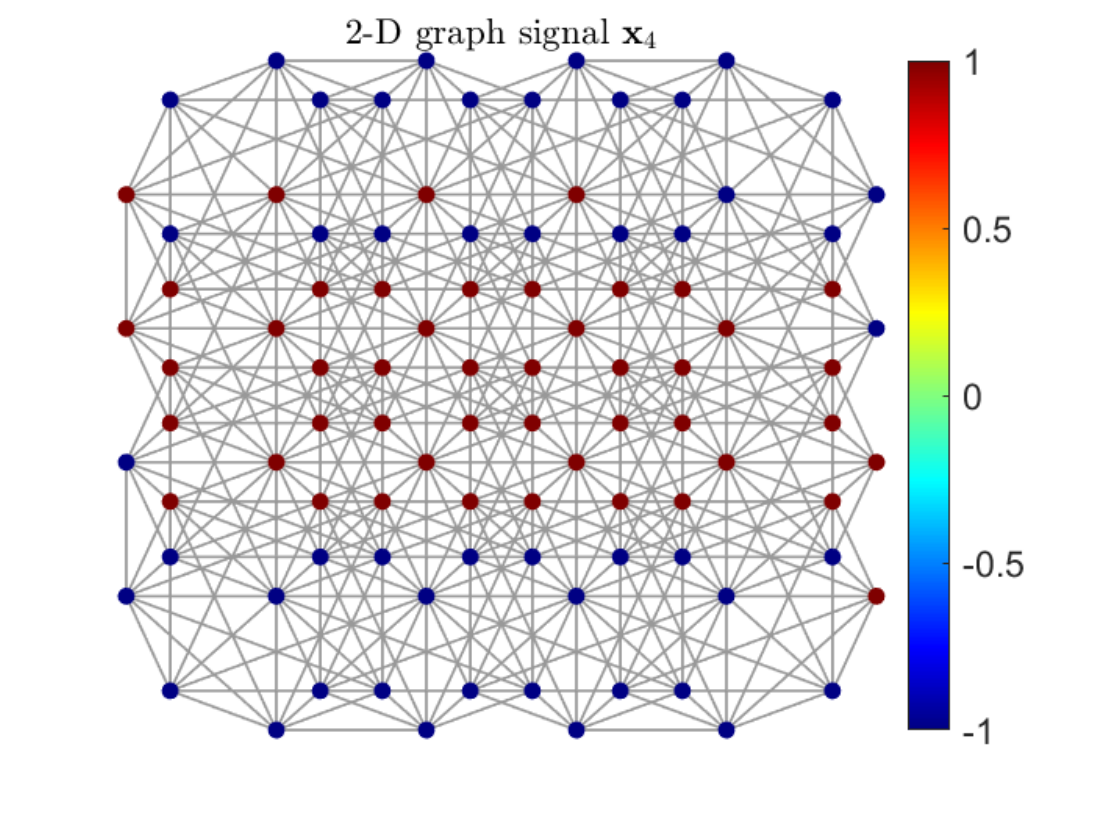} \label{4} 
}
\quad
\caption{2-D graph signals $\mathbf{x}_{1}$ to $\mathbf{x}_{4}$.}
\end{figure}

\begin{figure}
\subfigure[]{
	\includegraphics[scale=0.4]{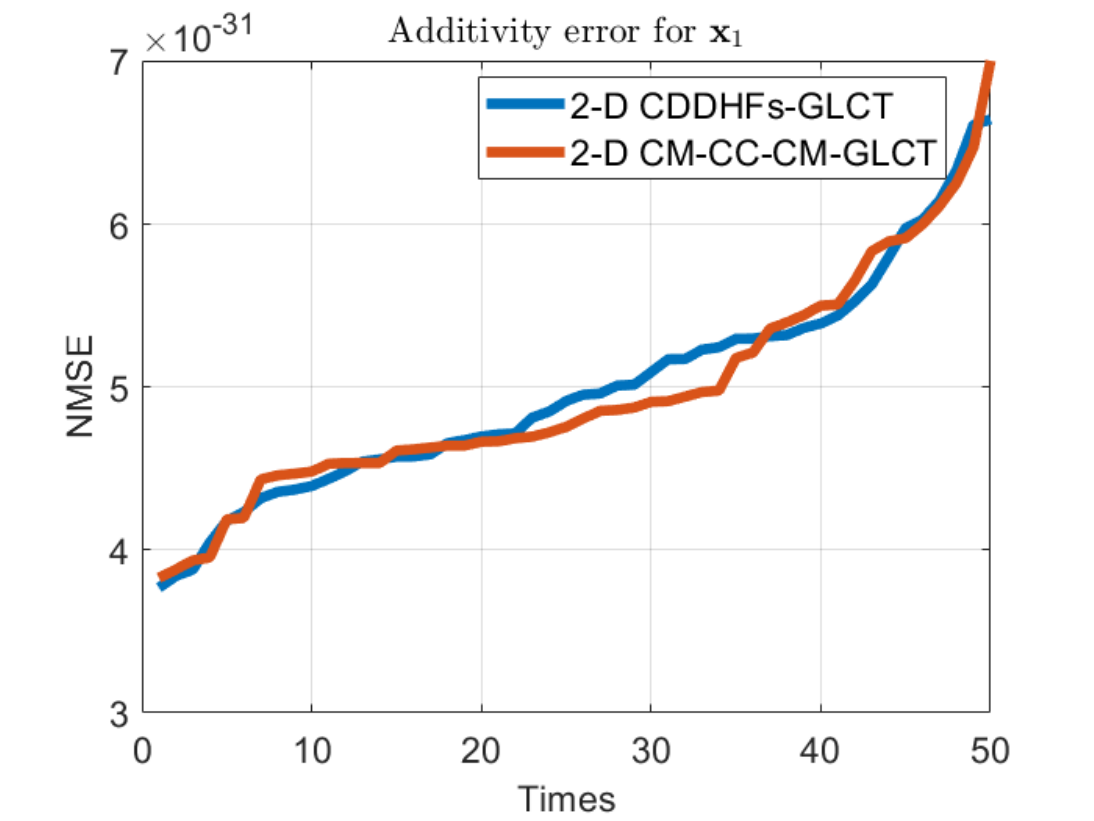} \label{5}
}
\quad
\subfigure[]{
	\includegraphics[scale=0.4]{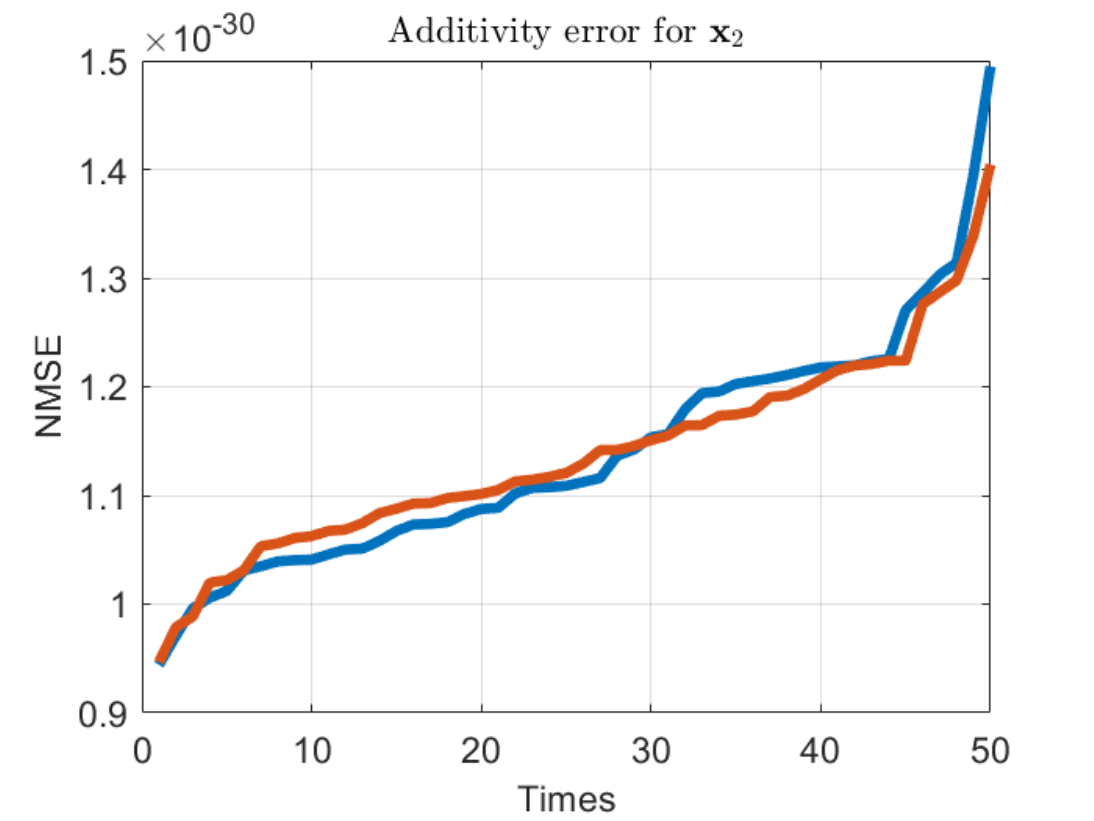} \label{6} 
}
\quad
	\subfigure[]{
		\includegraphics[scale=0.4]{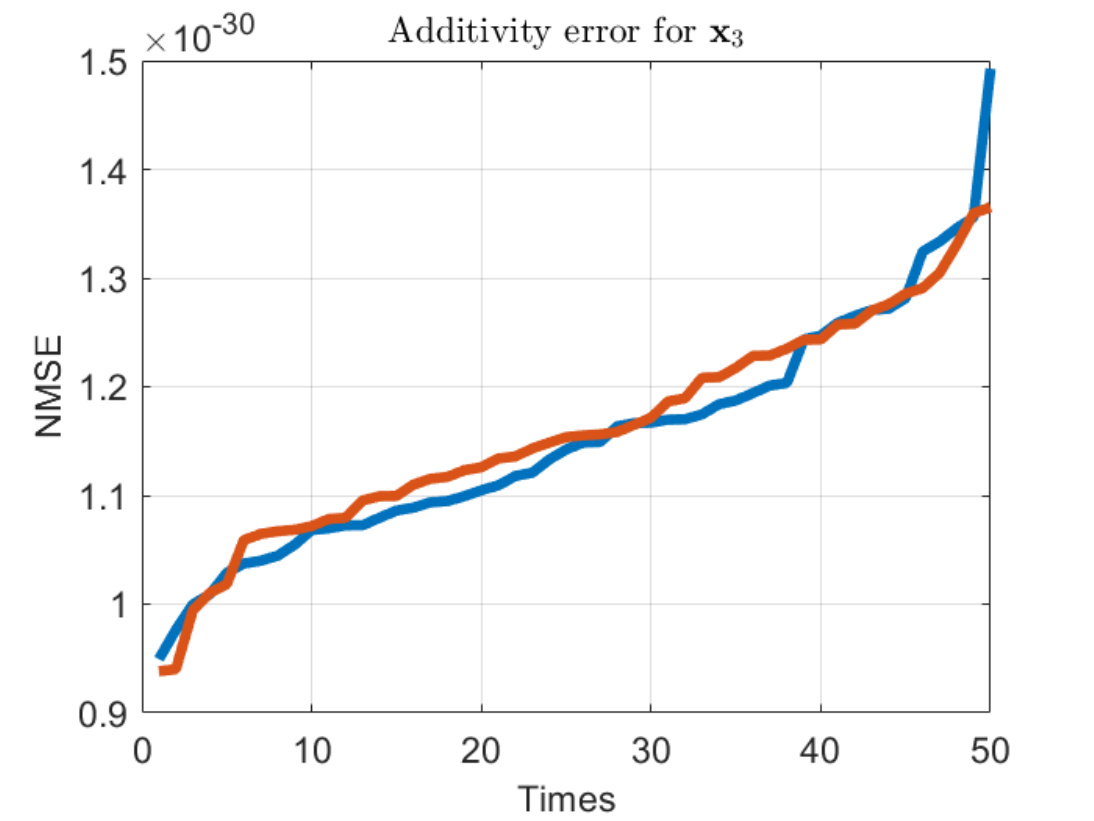} \label{1}
	}
	\quad
	\subfigure[]{
		\includegraphics[scale=0.4]{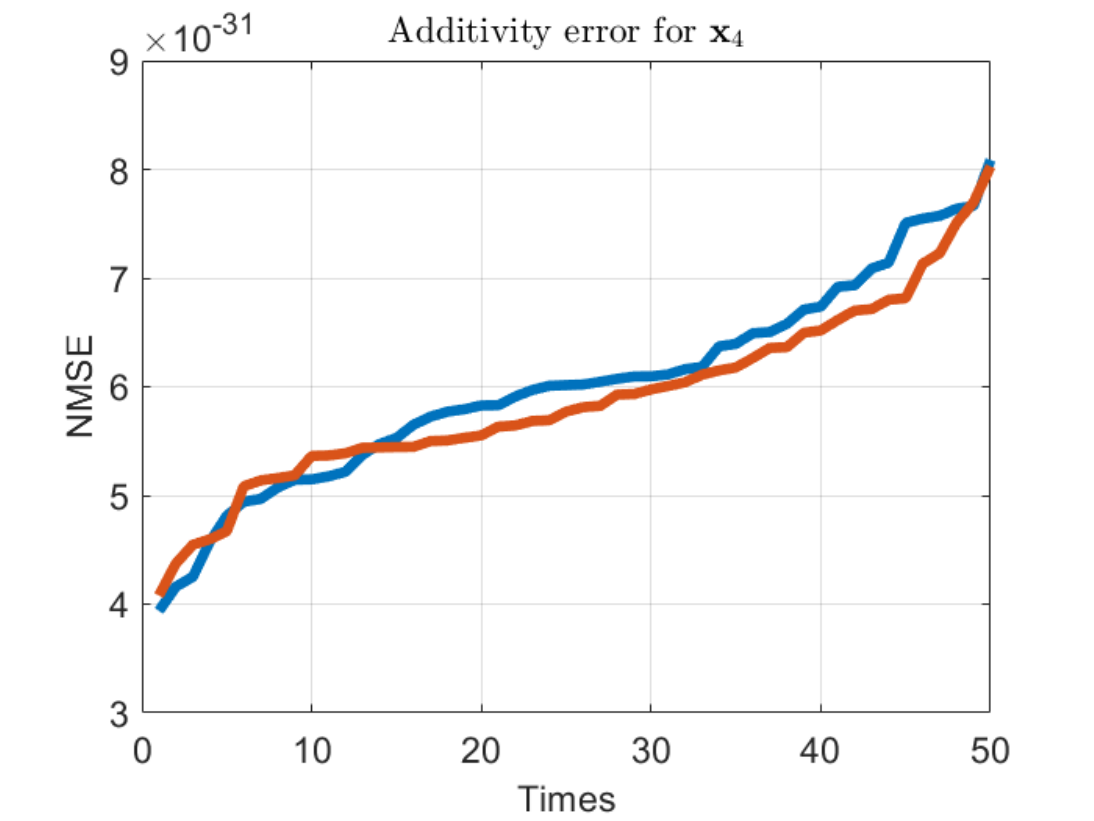} \label{1}
	}
	\quad
	\caption{Normalized mean-square errors (NMSEs) of the additivity property for 50 different sets of $\mathbf{M}_{1}$ and $\mathbf{M}_{2}$.}
\end{figure}

\begin{figure}
\subfigure[]{
	\includegraphics[scale=0.4]{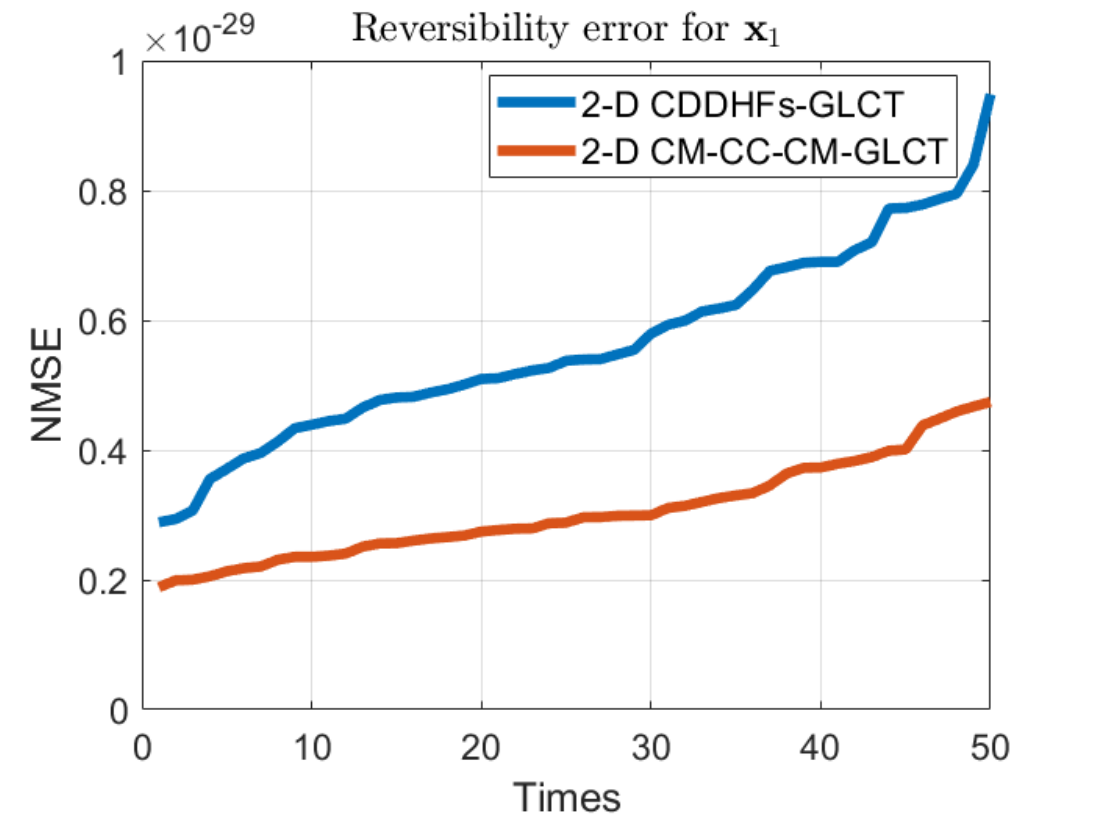} \label{1}
}
\quad
\subfigure[]{
	\includegraphics[scale=0.4]{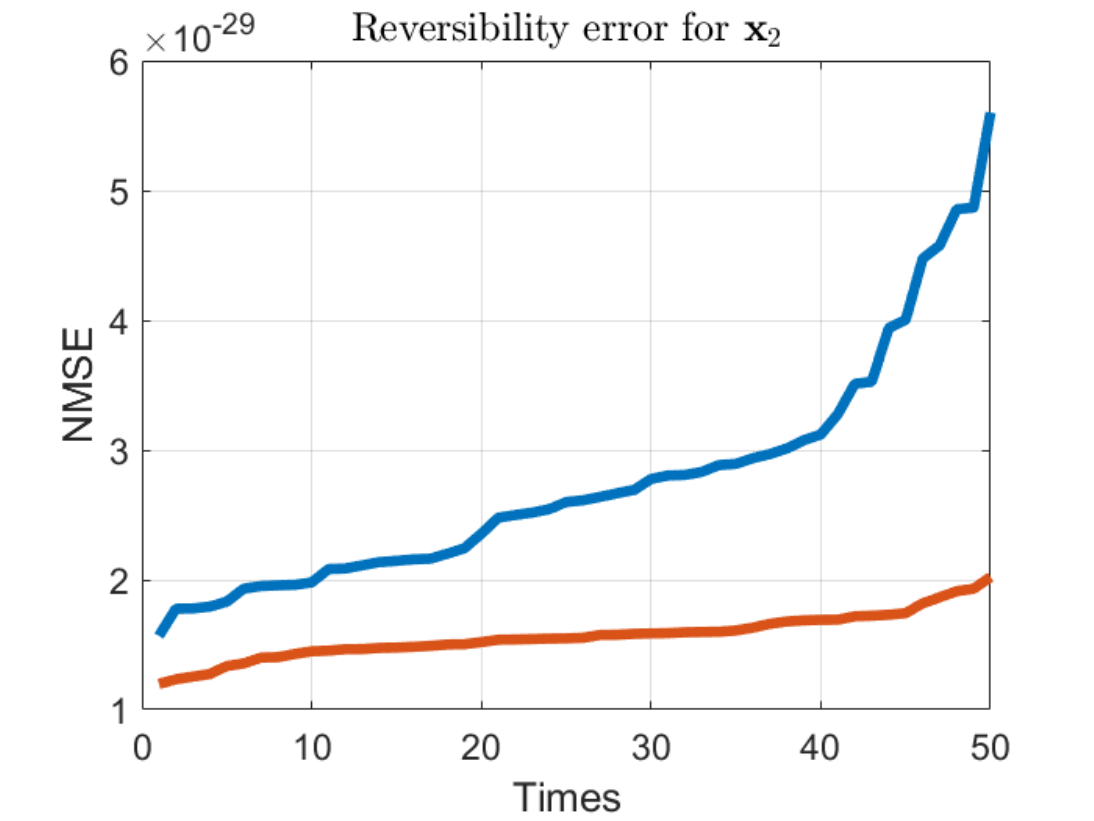} \label{2} 
}
\quad
\subfigure[]{
	\includegraphics[scale=0.4]{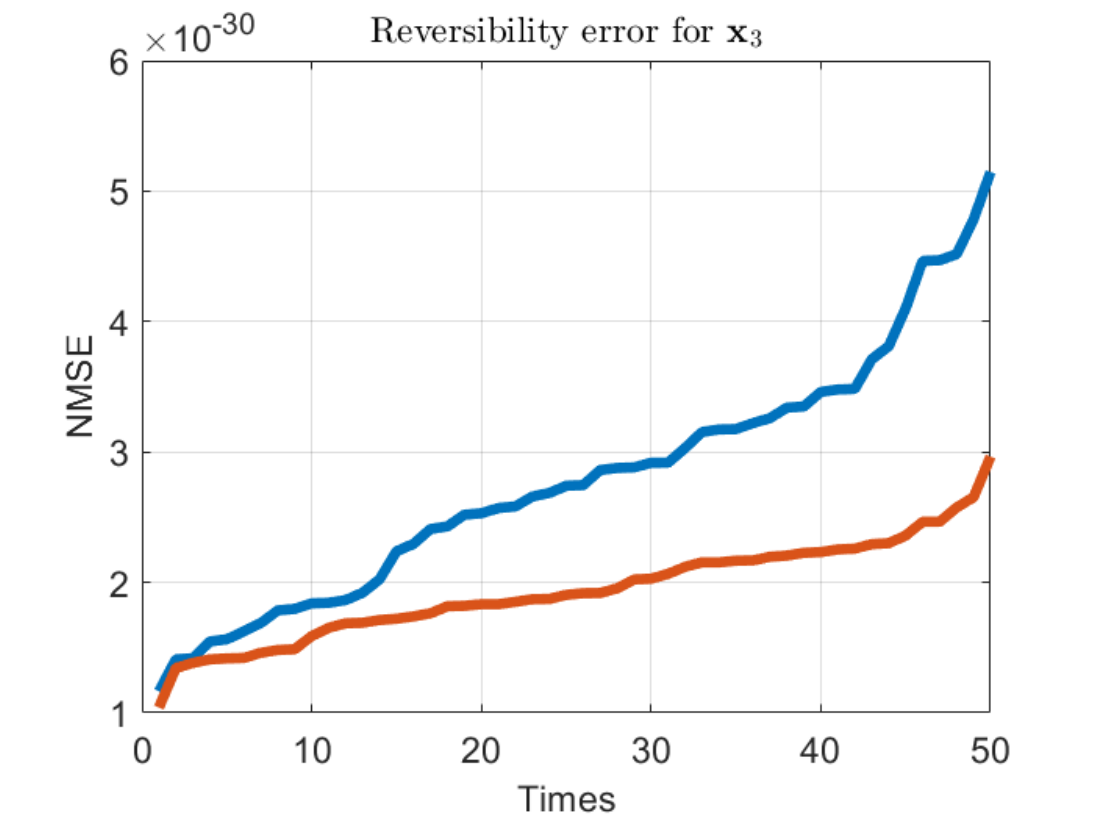} \label{3} 
}
\quad
\subfigure[]{
	\includegraphics[scale=0.4]{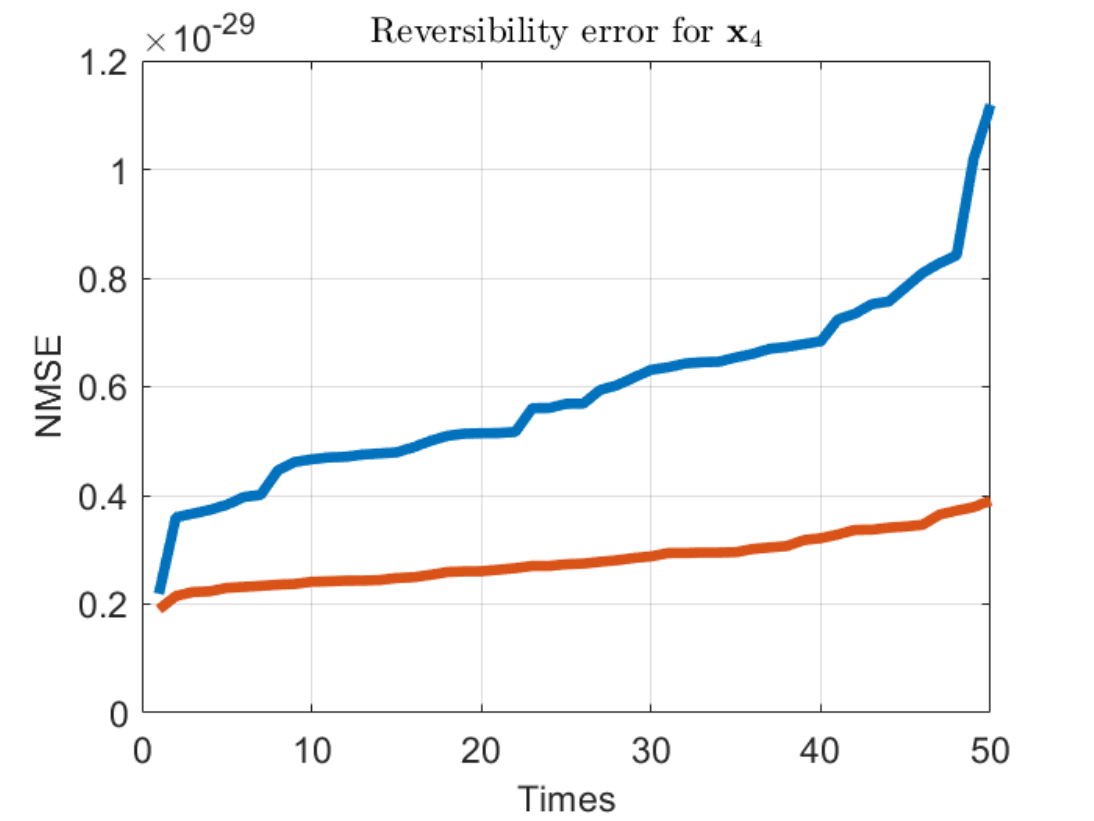} \label{4} 
}
\quad
\caption{Normalized mean-square errors (NMSEs) of the reversibility property for 50 different sets of $\mathbf{M}$.}
\end{figure}

\subsection{Reversibility property}
Next, we compare the reversibility of the 2-D CDDHFs-GLCTcand the 2-D CM-CC-CM-GLCT, by checking the NMSE of reversibility \cite{Pei2016FastDL}:
\begin{equation}
	{\rm NMSE}=\frac{\sum_{n=0}^{N_{1}N_{2}-1}\left | x\left [ n \right ] - O_{\rm{2-D \ GLCT}}^{\mathbf{ M}^{-1}}O_{\rm{2-D \ GLCT}}^{\mathbf{M}}\left ( x\left [ n \right ]  \right )    \right |^{2} }{\sum_{n=0}^{N_{1}N_{2}-1}\left | x\left [ n \right ]  \right |^{2}   } .
\end{equation}

The parameters in $\mathbf{M}$ are random numbers uniformly distributed on the interval [-2, 2], which are used for 50 simulation runs. The reversibility NMSE results in (44) is obtained. The NMSEs sorted in ascending order using $\mathbf{x}_{1}-\mathbf{x}_{4}$ as input signals are shown in Figure 3 respectively. As shown in figures, all NMSEs of 2-D CM-CC-CM-GLCT are below. In addition, the numerical verification of the average value of NMSEs for these graph signals for running 1000 times is shown in Table 3. If 2-D GLCT is reversible, there is no necessity to develop the 2-D IGLCT since it can be realized by the forward 2-D GLCT. Finally, Table 4 summarizes the comparison between the 2-D CDDHFs-GLCT and 2-D CM-CC-CM-GLCT.

\begin{table}[]
	\caption{Comparison of 1000 additivity running averages}
	\centering
	\begin{threeparttable}
	\begin{tabular}{|l|l|l|l|l|}
\hline
                & $\mathbf{x}_{1}$ & $\mathbf{x}_{2}$ & $\mathbf{x}_{3}$ & $\mathbf{x}_{4}$ \\ \hline
2-D CDDHFs-GLCT   & 4.9799           & 11.4810          & 3.9862           & 5.9608           \\ \hline
2-D CM-CC-CM-GLCT & 4.9398           & 11.4900          & 4.0500           & 5.9680           \\ \hline
\end{tabular}

	\begin{tablenotes}
	\footnotesize
	\item[*] The order of magnitude is $e^{-31}$ in the Table 2.
	\end{tablenotes}
\end{threeparttable}
\end{table}

\begin{table}[]
\caption{Comparison of 1000 reversibility running averages}
\centering
\begin{threeparttable}
		\begin{tabular}{|l|l|l|l|l|}
\hline
    
  & $\mathbf{x}_{1}$ & $\mathbf{x}_{2}$ & $\mathbf{x}_{3}$ & $\mathbf{x}_{4}$ \\ \hline
2-D CDDHFs-GLCT   & 5.5703           & 30.5880          & 2.6685           & 6.0682           \\ \hline
2-D CM-CC-CM-GLCT & 2.9121           & 15.1160          & 1.9557           & 2.8005           \\ \hline
\end{tabular}
	\begin{tablenotes}
	\footnotesize
	\item[*] The order of magnitude is $e^{-30}$ in the Table 3.
	\end{tablenotes}
\end{threeparttable}
\end{table}
\begin{table}[]
\caption{Comparison between the 2-D CDDHFs-GLCT and 2-D CM-CC-CM-GLCT}
\centering
	\begin{tabular}{|l|l|l|}
		\hline
		& 2-D CDDHFs-GLCT                                                             & 
		2-D CM-CC-CM-GLCT                                                      \\ \hline
		Complexity    & $O\left ( N_{1} ^{2}+N_{2} ^{2}  \right )$ & $O\left ( N_{1} \log_{2}{N}_{1}+ N_{2} \log_{2}{N}_{2}   \right )$ \\ \hline
		Additivity    & Approximate                                                               & Approximate                                                    \\ \hline
		Reversibility & Worse                                                                     & Better                                                         \\ \hline
	\end{tabular}
\end{table}

\section{Applications}
In various signal processing domains like storage, compression, and transmission, efficient signal representation is crucial. Some widely used techniques are based on extending signals to a suitable basis, and it is expected to capture most information about signals with few basis functions. To illustrate the effectiveness of this framework in M-D GLCT, we explore its application in data compression and compare it with M-D GFRFT to show the better compression effect of M-D GLCT. This section primarily focuses on the 2-D CM-CC-CM-GLCT.

We consider a 2-D graph $G$, which can be expressed as a Cartesian product graph of a ring graph $G_{1}\left ( N_{1}=100  \right )$ and a path graph $G_{2}\left ( N_{2}=15  \right )$, and the 2-D graph signal $\mathbf{x}$ is a random number evenly distributed in the interval $\left [ -10,10 \right ] $. The compression process is as follows. We use equation (26) to calculate the 2-D CM-CC-CM-GLCT of $\mathbf{x}$. Then rank the coefficients according to their absolute values. Let $0< \gamma < 1$, where $\gamma$ is the compression ratio. The maximum amplitude coefficient of the fraction $\gamma$ is reserved, and the remaining coefficient is set to zero. Finally, 2-D IGLCT is performed on the coefficients to obtain the compressed and reconstructed signal $\mathbf{x }_{com}  $. 

We choose three quantities to measure the compression loss \cite{Yan2021MultidimensionalGF}. The relative error (RE) is defined as
\begin{equation}
    {\rm{RE}}=\frac{\sum _{i_{1}\in G_{1}, i_{2}\in G_{2} }\left | x\left ( i_{1},i_{2} \right ) -x_{com}\left (  i_{1},i_{2} \right )   \right |  }{\sum _{i_{1}\in G_{1}, i_{2}\in G_{2} }\left | x\left (  i_{1},i_{2} \right )   \right | }
    .
\end{equation}

The normalized root mean square (NRMS) is defined by
\begin{equation}
    {\rm{NRMS}}=\frac{\sqrt{\sum _{i_{1}\in G_{1}, i_{2}\in G_{2} }\left ( x\left ( i_{1},i_{2} \right ) -x_{com}\left (  i_{1},i_{2} \right )\right )^{2}  }  }{\sqrt{\sum _{i_{1}\in G_{1}, i_{2}\in G_{2} }\left ( x\left ( i_{1},i_{2} \right ) -\bar{x} \right )^{2}  } },
    \end{equation}
where $\bar{x}=\frac{\sum _{i_{1}\in G_{1}, i_{2}\in G_{2}}x\left ( i_{1},i_{2} \right )}{N_{1}N_{2} }$ be the mean of the original signal. The correlation coefficient (CC) is defined by
\begin{equation}
    {\rm{CC}}=\frac{\sum _{i_{1}\in G_{1}, i_{2}\in G_{2}}\left ( x\left ( i_{1},i_{2} \right ) -\bar{x}\right )\left ( x_{com} \left ( i_{1},i_{2} \right ) -\bar{x}_{com} \right ) }{\sqrt{\sum _{i_{1}\in G_{1}, i_{2}\in G_{2}}\left ( x\left ( i_{1},i_{2} \right ) -\bar{x} \right )^{2} } \sqrt{\sum _{i_{1}\in G_{1}, i_{2}\in G_{2}}\left ( x_{com} \left ( i_{1},i_{2} \right ) -\bar{x}_{com} \right )^{2} }  } .
\end{equation}

RE and NRMS are proposed to measure the error between the compressed signal and the original signal. In both cases, smaller error means better compression. CC measures compression efficiency, where a larger value means that the compressed signal contains more information from the original signal. We employ the definition of 2-D GFRFT in (10) to conduct experiments across different compression ratios $\gamma$, with results depicted in Figures 4. The results show that the optimal rotation angle parameter $\alpha$ is 1 under different compression ratios, degenerating into 2-D GFT. Using the definition of 2-D CM-CC-CM-GLCT in (26), various experiments are carried out for different compression ratios $\gamma$, in order to find a set of parameters to make the results better than the optimal index under the framework of 2-D GFRFT. As shown in Figure 5. Specific parameter indices are detailed in Table 5-7. Obviously, compared to 2-D GFRFT results, 2-D CM-CC-CM-GLCT provides a better transformation basis and produces smaller errors. Compared with M-D GFRFT, M-D CM-CC-CM-GLCT can be closer to the original data under the same compression ratio, that is, in smaller differences between compressed and original data. This shows that M-D CM-CC-CM-GLCT can effectively preserve more data details and structural information in the compression process. Consequently, M-D CM-CC-CM-GLCT can more accurately restore the original data characteristics without significant distortion or errors introduced by compression. Parameter selection in most experiments involves iterative refinement, aided by adjustments in linear regularization parameters to achieve better compression.

\begin{figure}
\subfigure[]{
    \includegraphics[scale=0.4]{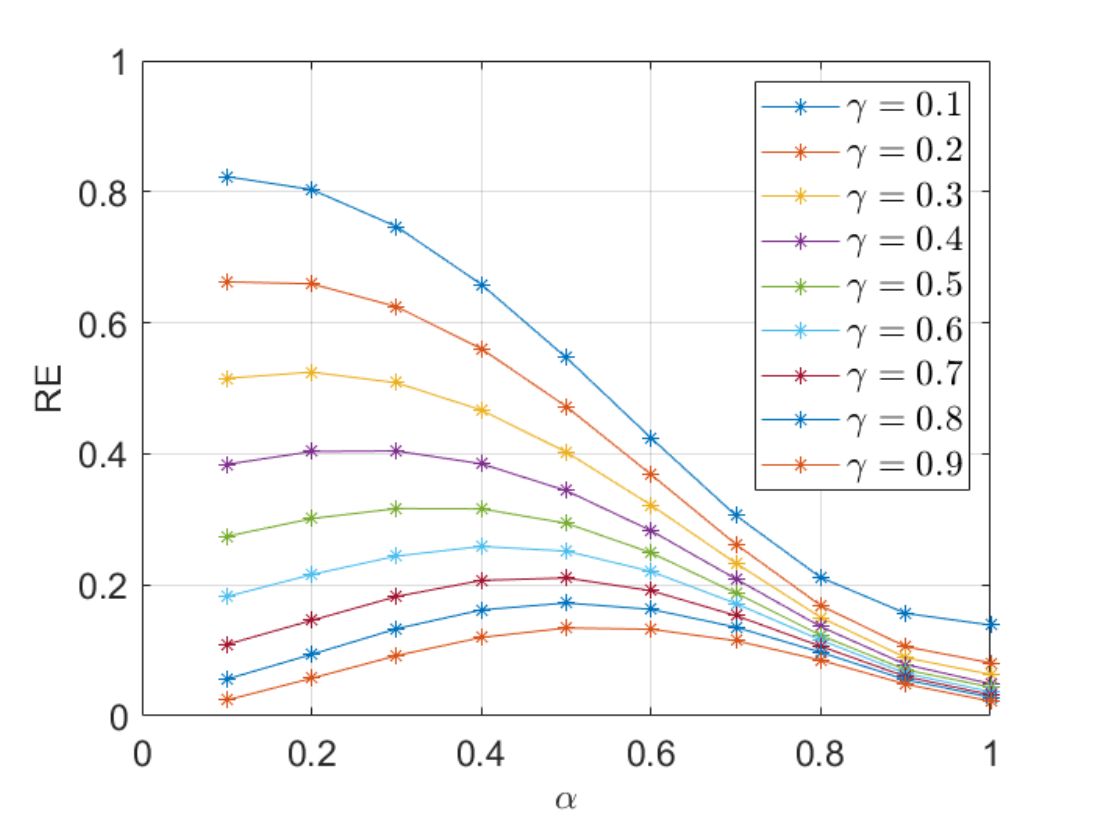} \label{1}
}
\quad
\subfigure[]{
    \includegraphics[scale=0.4]{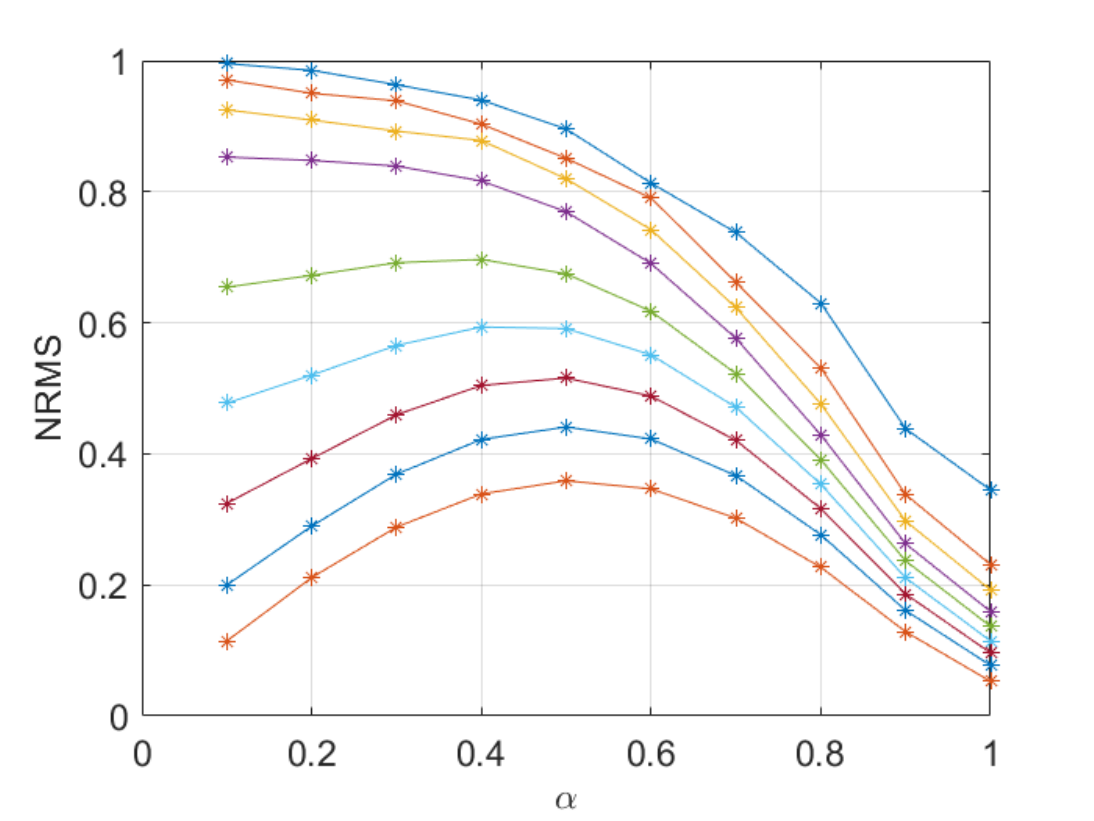} \label{2} 
}
\quad
\begin{center}
\subfigure[]{
    \includegraphics[scale=0.4]{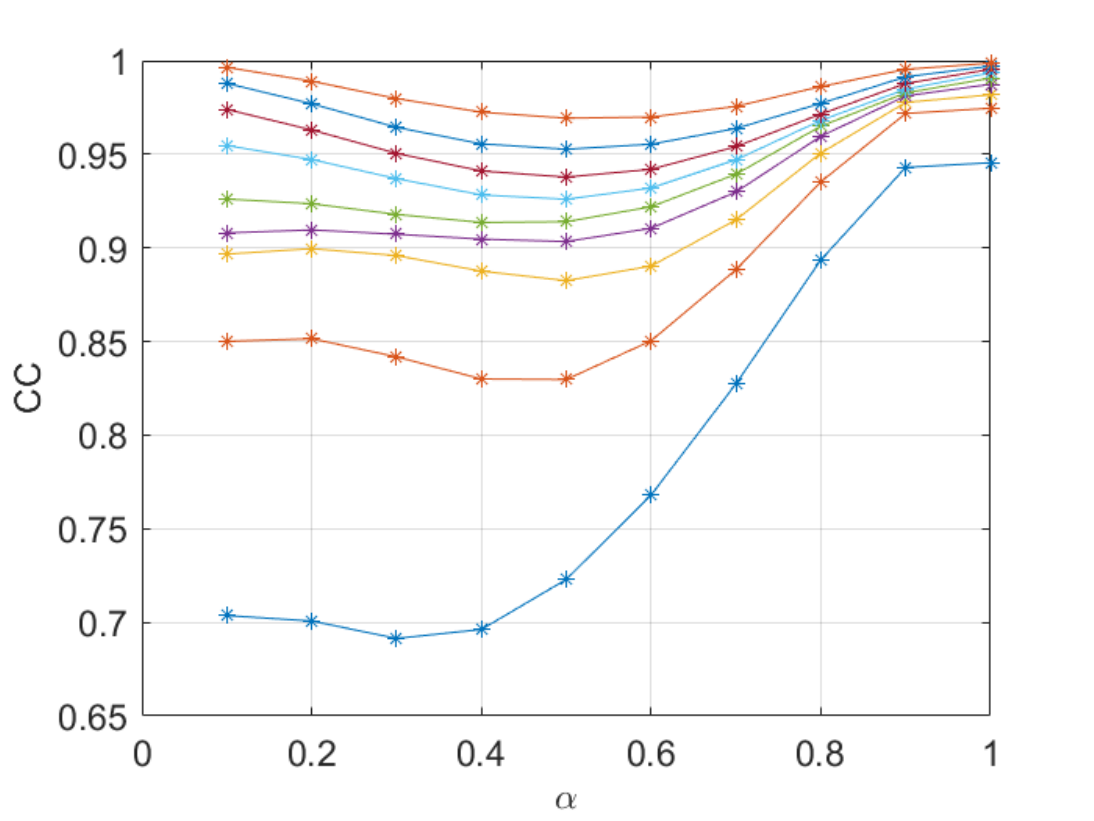} \label{3} 
}
\end{center}
\quad
\caption{RE, NRMS and CC of 2-D GFRFT with different rotation angles $\alpha$ for different compression parameters $\gamma$.}
\end{figure}
\begin{figure}
\subfigure[]{
    \includegraphics[scale=0.4]{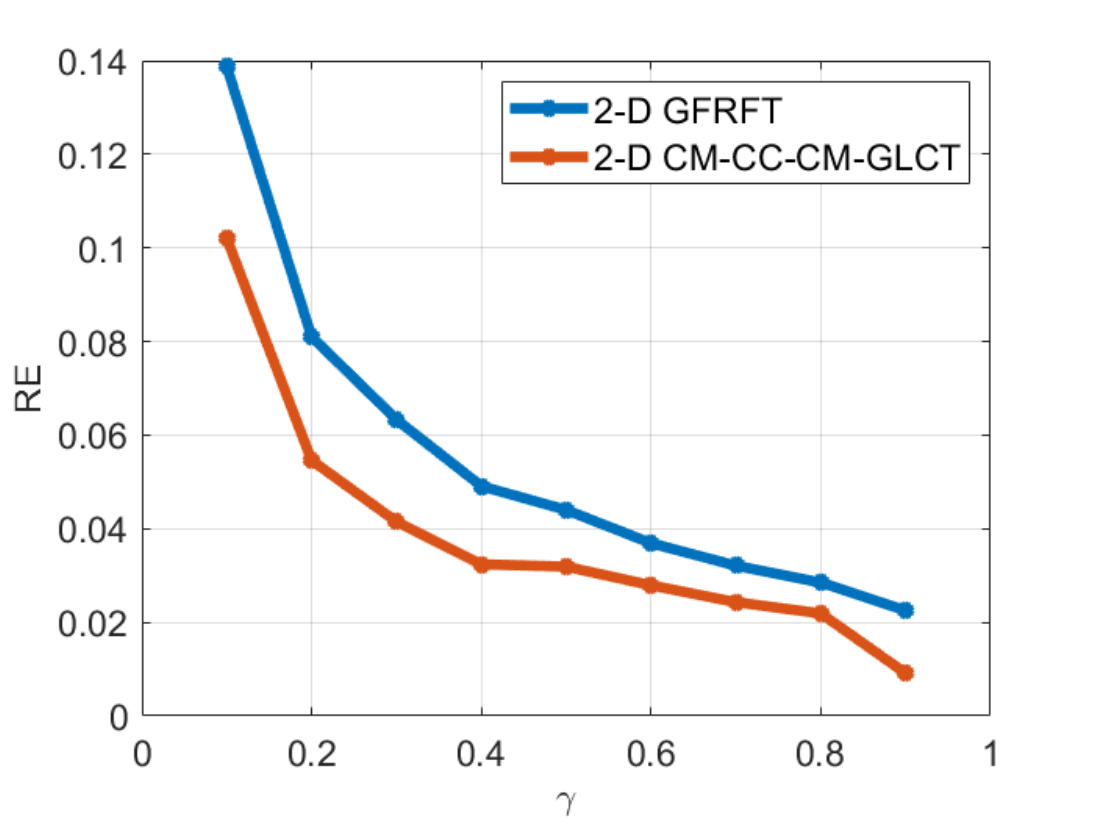} \label{1}
}
\quad
\subfigure[]{
    \includegraphics[scale=0.4]{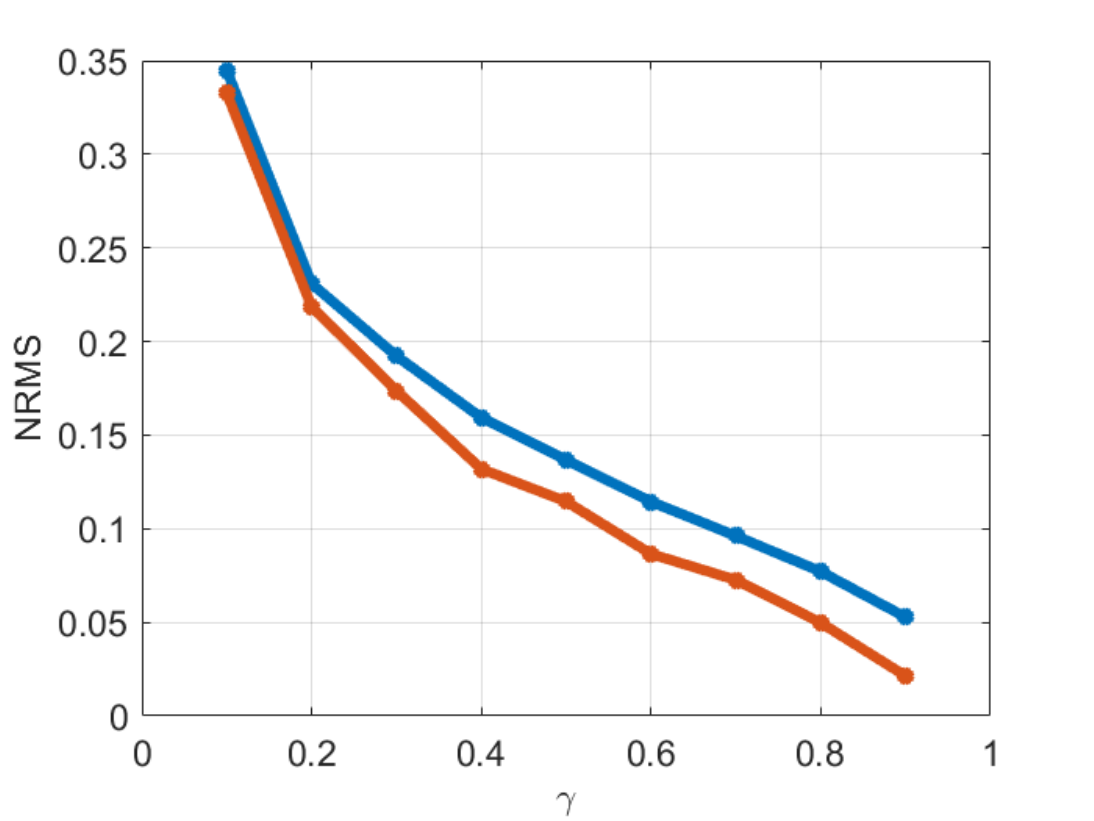} \label{2} 
}
\quad
\begin{center}
\subfigure[]{
    \includegraphics[scale=0.4]{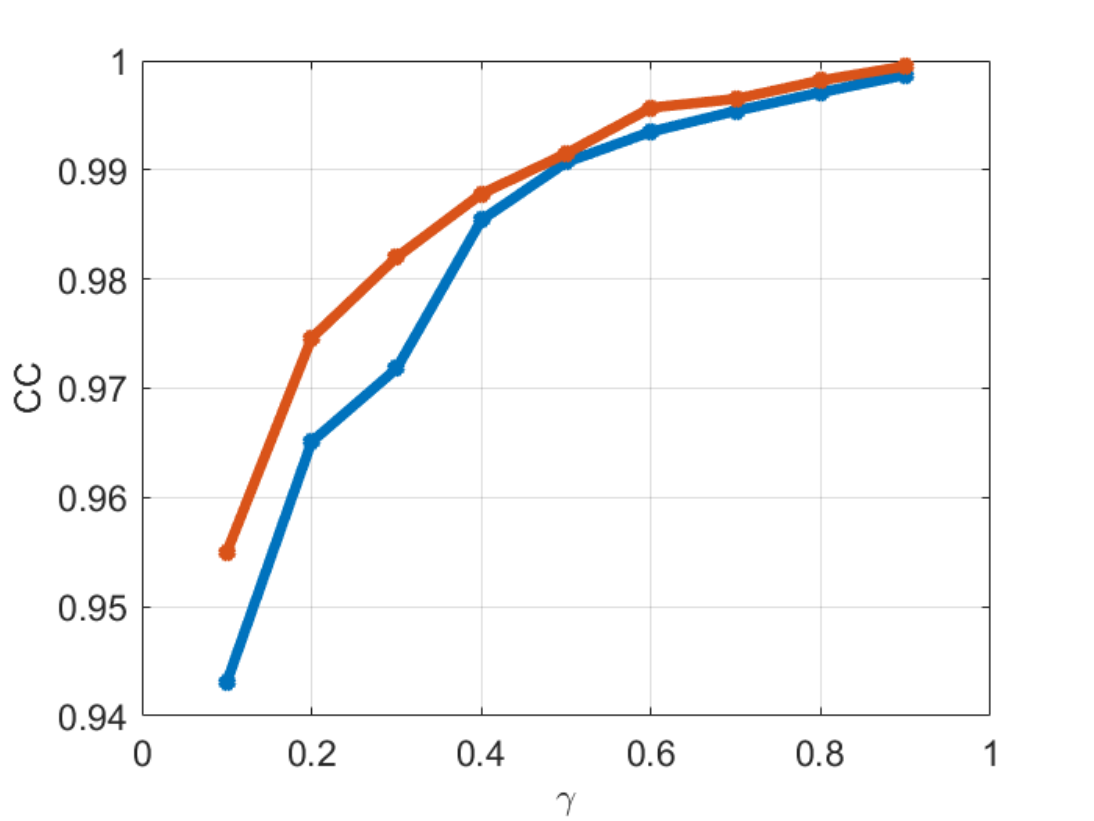} \label{3} 
}
\end{center}
\quad
\caption{Comparison between the 2-D GFRFT and the 2-D CM-CC-CM-GLCT with different compression parameters $\gamma$.}
\end{figure}

\begin{table}[]
\caption{RE of 2-D GFRFT and 2-D CM-CC-CM-GLCT with different $\gamma$}
	\centering
	\resizebox{\textwidth}{!}{
 \begin{threeparttable}
\begin{tabular}{|l|l|l|l|l|l|l|l|l|l|}
\hline
RE                & $\gamma =0.1$ & $\gamma =0.2$ & $\gamma =0.3$ & $\gamma =0.4$ & $\gamma =0.5$ & $\gamma =0.6$ & $\gamma =0.7$ & $\gamma =0.8$ & $\gamma =0.9$ \\ \hline
2-D GFRFT         & 13.87\%       & 8.10\%        & 6.33\%        & 4.91\%        & 4.40\%        & 3.69\%        & 3.21\%        & 2.85\%        & 2.25\%        \\ \hline
2-D CM-CC-CM-GLCT & 10.20\%       & 5.46\%        & 4.15\%        & 3.24\%        & 3.19\%        & 2.79\%        & 2.43\%        & 2.19\%        & 0.91\%        \\ \hline
\end{tabular}

\begin{tablenotes}
	\footnotesize
	\item[*] In Table 5, for different compression ratios $\gamma $, the optimal rotation angle $\alpha $ is selected for 2-D GFRFT, where $\alpha $=1. The parameters selected for 2-D CM-CC-CM-GLCT are as follows: a=0.10, b=0.60, c=-0.20, d=8.80 for $\gamma $=0.1; a=0.40, b=0.10, c=-0.60, d=2.35 for $\gamma $=0.2; a=0.10, b=-0.20, c=0.90, d=8.20 for $\gamma $=0.3; a=0.30, b=-0.20, c=-1.40, d=4.27 for $\gamma $=0.4; a=0.80, b=-1.20, c=1.40, d=-0.85 for $\gamma $=0.5; a=1.20, b=1.40, c=-0.30, d=0.48 for $\gamma $=0.6; a=-1.30, b=0.50, c=0.40, d=-0.92 for $\gamma $=0.7; a=-0.90, b=0.30, c=-1.20, d=-0.71 for $\gamma $=0.8; a=0.40, b=-1.10, c=0.70, d=0.58 for $\gamma $=0.9.
	\end{tablenotes}
 
 \end{threeparttable}
 }
\end{table}

\begin{table}[]
\caption{NRMS of 2-D GFRFT and 2-D CM-CC-CM-GLCT with different $\gamma$}
	\centering
	\resizebox{\textwidth}{!}{
  \begin{threeparttable}
\begin{tabular}{|l|l|l|l|l|l|l|l|l|l|}
\hline
NRMS                & $\gamma =0.1$ & $\gamma =0.2$ & $\gamma =0.3$ & $\gamma =0.4$ & $\gamma =0.5$ & $\gamma =0.6$ & $\gamma =0.7$ & $\gamma =0.8$ & $\gamma =0.9$ \\ \hline
2-D GFRFT         & 34.47\%       & 23.09\%        & 19.22\%        & 15.93\%        & 13.67\%        & 11.44\%        & 9.61\%        & 7.71\%        & 5.27\%        \\ \hline
2-D CM-CC-CM-GLCT & 33.25\%       & 21.86\%        & 17.37\%        & 13.19\%        & 11.45\%        & 8.65\%        & 7.26\%        & 4.97\%        & 2.13\%        \\ \hline
\end{tabular}
\begin{tablenotes}
	\footnotesize
	\item[*] In Table 6, for different compression ratios $\gamma $, the optimal rotation angle $\alpha $ is selected for 2-D GFRFT, where $\alpha $=1. The parameters selected for 2-D CM-CC-CM-GLCT are as follows: a=-1.80, b=0.30, c=-1.20, d=-0.36 for $\gamma $=0.1; a=-0.40, b=0.30, c=1.50, d=-3.60 for $\gamma $=0.2; a=0.30, b=0.10, c=-0.60, d=3.53 for $\gamma $=0.3; a=0.40, b=-1.00, c=0.70, d=0.75 for $\gamma $=0.4; a=1.30, b=-0.20, c=1.40, d=0.55 for $\gamma $=0.5; a=1.20, b=0.50, c=-1.00, d=0.42 for $\gamma $=0.6; a=1.50, b=-1.40, c=0.20, d=0.48 for $\gamma $=0.7; a=-0.70, b=0.30, c=1.50, d=-2.07 for $\gamma $=0.8; a=0.30, b=-0.20, c=1.80, d=2.13 for $\gamma $=0.9.
	\end{tablenotes}
 
 \end{threeparttable}
}
\end{table}
\begin{table}[]
\caption{CC of 2-D GFRFT and 2-D CM-CC-CM-GLCT with different $\gamma$}
	\centering
	\resizebox{\textwidth}{!}{
 \begin{threeparttable}
\begin{tabular}{|l|l|l|l|l|l|l|l|l|l|}
\hline
CC                & $\gamma =0.1$ & $\gamma =0.2$ & $\gamma =0.3$ & $\gamma =0.4$ & $\gamma =0.5$ & $\gamma =0.6$ & $\gamma =0.7$ & $\gamma =0.8$ & $\gamma =0.9$ \\ \hline
2-D GFRFT         & 94.31\%       & 96.51\%        & 97.18\%        & 98.54\%        & 99.07\%        & 99.35\%        & 99.54\%        & 99.71\%        & 99.87\%        \\ \hline
2-D CM-CC-CM-GLCT & 95.50\%       & 97.46\%        & 98.20\%        & 98.78\%        & 99.15\%        & 99.57\%        & 99.65\%        & 99.82\%        & 99.95\%        \\ \hline
\end{tabular}
\begin{tablenotes}
	\footnotesize
	\item[*] In Table 7, for different compression ratios $\gamma $, the optimal rotation angle $\alpha $ is selected for 2-D GFRFT, where $\alpha $=1. The parameters selected for 2-D CM-CC-CM-GLCT are as follows: a=-0.20, b=0.70, c=0.40, d=-6.40 for $\gamma $=0.1; a=-0.20, b=0.50, c=1.10, d=-7.75 for $\gamma $=0.2; a=0.30, b=0.40, c=-0.80, d=2.27 for $\gamma $=0.3; a=1.40, b=0.10, c=-0.60, d=0.67 for $\gamma $=0.4; a=0.90, b=-1.10, c=0.40, d=0.62 for $\gamma $=0.5; a=-0.60, b=1.40, c=0.90, d=-3.77 for $\gamma $=0.6; a=0.70, b=1.80, c=-0.50, d=0.14 for $\gamma $=0.7; a=0.40, b=0.10, c=-0.60, d=2.35 for $\gamma $=0.8; a=0.10, b=0.50, c=1.20, d=16.00 for $\gamma $=0.9.
	\end{tablenotes}
 \end{threeparttable}
}
\end{table}

\section{Conclusions}
In this paper, a multi-dimensional graph signal processing method is proposed, which preserves the dimension information of multi-dimensional graphs. We introduce two-dimensional graph linear canonical transforms, namely, 2-D CDDHFs-FLCT and 2-D CM-CC-CM-GLCT, derived from CDDHFs decomposition and CM-CC-CM decomposition. Subsequently, we extend these transforms to multi-dimensional, resulting in M-D CDDHFs-GLCT and M-D CM-CC-CM-GLCT, respectively. Considering computational complexity, additivity and reversibility, M-D CM-CC-CM-GLCT is compared with M-D CDDHFs-GLCT. Theoretical analysis shows that the computational complexity of M-D CM-CC-CM-GLCT algorithm is obviously reduced. Simulation results indicate that M-D CM-CC-CM-GLCT achieves comparable additivity to M-D CDDHFs-GLCT, while M-D CM-CC-CM-GLCT exhibits better reversibility. Finally, we apply M-D GLCT to data compression to illustrate its application advantages. Experiments show that compared with M-D GFRFT, M-D GLCT achieves closer fidelity to the original data at the same compression ratio, that is, in smaller differences between the compressed and original data. This reflects the superiority of M-D GLCT in the algorithm design and implementation for data compression.



\bibliographystyle{elsarticle-num}  
\bibliography{ref}  

\begin{thebibliography}{10}
\expandafter\ifx\csname url\endcsname\relax
  \def\url#1{\texttt{#1}}\fi
\expandafter\ifx\csname urlprefix\endcsname\relax\def\urlprefix{URL }\fi
\expandafter\ifx\csname href\endcsname\relax
  \def\href#1#2{#2} \def\path#1{#1}\fi

\bibitem{Shuman2012TheEF}
D.~I. Shuman, S.~K. Narang, P.~Frossard, A.~Ortega, P.~Vandergheynst, The emerging field of signal processing on graphs: extending high-dimensional data analysis to networks and other irregular domains, IEEE Signal Processing Magazine, 30 (2012) 83--98.

\bibitem{Sandryhaila2012DiscreteSP}
A.~Sandryhaila, J.~M.~F. Moura, Discrete signal processing on graphs, IEEE Transactions on Signal Processing, 61 (2012) 1644--1656.

\bibitem{Sandryhaila2014BigDA}
A.~Sandryhaila, J.~M.~F. Moura, Big data analysis with signal processing on graphs: representation and processing of massive data sets with irregular structure, IEEE Signal Processing Magazine, 31 (2014) 80--90.

\bibitem{Alikaifolu2024WienerFI}
T.~Alikaşifoğlu, B.~Kartal, A.~Koç, Wiener filtering in joint time-vertex fractional {F}ourier domains, IEEE Signal Processing Letters, 31 (2024) 1319--1323.

\bibitem{Domingos2020GraphFT}
J.~Domingos, J.~M.~F. Moura, Graph {F}ourier transform: a stable approximation, IEEE Transactions on Signal Processing, 68 (2020) 4422--4437.

\bibitem{Yang2019GraphFT}
L.~Yang, A.~Qi, C.~Huang, J.~Huang, Graph {F}ourier transform based on $l_{1}$ norm variation minimization, Applied and Computational Harmonic Analysis, 52 (2021) 348--365.

\bibitem{Jiang2019NonsubsampledGF}
J.~Jiang, C.~Cheng, Q.~Sun, Nonsubsampled graph filter banks: theory and distributed algorithms, IEEE Transactions on Signal Processing, 67 (2019) 3938--3953.

\bibitem{Ramakrishna2020AUG}
R.~Ramakrishna, H.-T. Wai, A.~Scaglione, A user guide to low-pass graph signal processing and its applications: tools and applications, IEEE Signal Processing Magazine, 37 (2020) 74--85.

\bibitem{Xiao2021DistributedNP}
Z.~Xiao, H.~Fang, X.~Wang, Distributed nonlinear polynomial graph filter and its output graph spectrum: filter analysis and design, IEEE Transactions on Signal Processing, 69 (2021) 1--15.

\bibitem{Kurokawa2017MultidimensionalGF}
T.~Kurokawa, T.~Oki, H.~Nagao, Multi-dimensional graph {F}ourier transform, ArXiv, abs/1712.07811 (2017).

\bibitem{Natali2020ForecastingMP}
A.~Natali, E.~Isufi, G.~Leus, Forecasting multi-dimensional processes over graphs, ICASSP 2020 IEEE International Conference on Acoustics, Speech and Signal Processing (ICASSP), (2020) 5575--5579.

\bibitem{Varma2018SAMPLINGTF}
R.~Varma, J.~Kovacevic, Sampling theory for graph signals on product graphs, 2018 IEEE Global Conference on Signal and Information Processing (GlobalSIP), (2018) 768--772.

\bibitem{Wang2017TheFF}
Y.~Wang, B.~Li, Q.~Cheng, The fractional {F}ourier transform on graphs, 2017 Asia-Pacific Signal and Information Processing Association Annual Summit and Conference (APSIPA ASC), (2017) 105--110.

\bibitem{Wu2019FractionalSG}
J.~Wu, F.~Wu, Q.~Yang, Y.~Kong, X.~Liu, Y.~Zhang, L.~Senhadji, H.~Shu, Fractional spectral graph wavelets and their applications, ArXiv, abs/1902.10471 (2019).

\bibitem{Yan2021WindowedFF}
F.~Yan, B.~Li, Windowed fractional {F}ourier transform on graphs: properties and fast algorithm, Digital Signal Processing, 118 (2021) 103210.

\bibitem{Kartal2022JointTF}
B.~Kartal, E.~{\"O}zg{\"u}nay, A.~Koç, Joint time-vertex fractional {F}ourier transform, ArXiv, abs/2203.07655 (2022).

\bibitem{Yan2021MultidimensionalGF}
F.~Yan, B.~Li, Multi-dimensional graph fractional {F}ourier transform and its application to data compression, Digital Signal Processing, 129 (2021) 103683.

\bibitem{Wei2021SparseDL}
D.~Wei, H.~Hu, Sparse discrete linear canonical transform and its applications, Signal Processing, 183 (2021) 108046.

\bibitem{Urynbassarova2021DiscreteQL}
D.~Urynbassarova, A.~A. Teali, F.~Zhang, Discrete quaternion linear canonical transform, Digital Signal Processing, 122 (2021) 103361.

\bibitem{Goel2022ApplicationsOT}
N.~Goel, S.~Gabarda, Applications of the linear canonical transform to digital image processing, Journal of the Optical Society of America A-Optics Image Science and Vision, 39 9 (2022) 1729--1738.

\bibitem{Ciobanu2021ModelingCC}
A.~A. Ciobanu, D.~D. Brown, P.~J. Veitch, D.~J. Ottaway, Modeling circulating cavity fields using the discrete linear canonical transform, Journal of the Optical Society of America A-Optics Image Science and Vision, 38 9 (2021) 1293--1303.

\bibitem{Zhang2022DiscreteLC}
Y.~Zhang, B.~Li, Discrete linear canonical transform on graphs, Digital Signal Processing, 135 (2022) 103934.

\bibitem{Pei2002EigenfunctionsOL}
S.~C. Pei, J.~Ding, Eigenfunctions of linear canonical transform, IEEE Transactions on Signal Processing, 50 (2002) 11--26.

\bibitem{Satyan2010ChirpMB}
N.~Satyan, G.~Rakuljic, A.~Yariv, Chirp multiplication by four wave mixing for wideband swept-frequency sources for high resolution imaging, Journal of Lightwave Technology, 28 (2010) 2077--2083.

\bibitem{ZhangGraphLC}
N.~Li, Z.~Zhang, J.~Han, Y.~Chen, C.~Cao, Graph linear canonical transform based on {CM-CC-CM} decomposition, submitted.

\bibitem{Imrich2008TopicsIG}
W.~Imrich, S.~Klavar, D.~F. Rall, Topics in graph theory: graphs and their cartesian product (2008).

\bibitem{Kadambari2019LearningPG}
S.~K. Kadambari, S.~P. Chepuri, Learning product graphs from multidomain signals, ICASSP 2020 IEEE International Conference on Acoustics, Speech and Signal Processing (ICASSP), (2020) 5665--5669.

\bibitem{Ko2010FastAA}
A.~Koç, H.~M. Ozaktas, L.~Hesselink, Fast and accurate computation of two-dimensional non-separable quadratic-phase integrals, Journal of the Optical Society of America A-Optics Image Science and Vision, 27 6 (2010) 1288--302.

\bibitem{delaCruz2021OnTI}
R.~J.~D. Cruz, E.~N. Reyes, On the {I}wasawa decomposition of a perplectic matrix, Communications in Algebra, 49 (2021) 932--947.

\bibitem{Pei2016FastDL}
S.~C. Pei, S.~G. Huang, Fast discrete linear canonical transform based on {CM-CC-CM} decomposition and {FFT}, IEEE Transactions on Signal Processing, 64 (2016) 855--866.

\end{thebibliography}
\end{document}